\documentclass{article}

\usepackage{lineno}

\usepackage{booktabs}
\usepackage{amsmath}
\usepackage{authblk}
\usepackage{array,ragged2e,xcolor}
\usepackage{multirow}
\usepackage{graphicx}
\usepackage[numbers]{natbib}
\usepackage{hyperref}
\usepackage{fixltx2e}
\usepackage{subcaption}

\usepackage[preprint]{tackling_climate_workshop_style}

\usepackage[utf8]{inputenc} 
\usepackage[T1]{fontenc}    
\usepackage{hyperref}       
\usepackage{url}            
\usepackage{booktabs}       
\usepackage{amsfonts}       
\usepackage{nicefrac}       
\usepackage{microtype}      

\title{\textit{} Improving Tropical Cyclone Forecasts through Objective Ensemble Track-Error Reduction}

\author[1]{\textbf{Shuochen Wang}}
\author[2]{\textbf{Mark S. Allen}}
\author[1, 3]{\textbf{Auroop R. Ganguly}}
\affil[1]{Sustainability and Data Sciences Laboratory, Northeastern University}
\affil[2]{InquisIT}
\affil[3]{The Institute for Experiential AI, Northeastern University}

\begin{document}

\maketitle

\begin{abstract}

Accurate tropical cyclone (TC) track guidance is essential for warnings, evacuation planning, and risk communication, but outlying members and distinct track scenarios can degrade raw ensemble forecasts. Statistical and machine-learning (ML) methods proposed to improve ensemble guidance are difficult to compare because previous studies often use different forecast systems, basins, periods, and verification methods. Using a common verification framework, this study compares six objective methods (four statistical trimming and two ML methods) with the full ensemble mean for 138 TCs forecast by GEFS, MOGREPS-G, and IFS during 2016--2024. Dynamic trimming (DT) is the most consistent statistical method, reducing mean total track error by 8.0\%, 13.4\%, and 16.1\% at short, medium, and long ranges. Its largest benefits occur in high-spread cases, where it can reach 70\% of the low-error bound. Short-lead-error trimming (SLET) provides useful delayed guidance after early observations become available, but its improvement decreases with lead time. Regression-mixture clustering (RC) shows potential in low-spread cases and provides the largest improvement over the control run. The gradient boosted tree (GBT) produces the largest medium-range improvement and exceeds the theoretical trimming bound, suggesting strong potential, particularly when the observed track is poorly represented by the ensemble family. However, its 12.8\% short-range degradation and gains concentrated in a small subset of cases indicate overcorrection when the ensemble mean is already accurate, highlighting the need for additional constraints. These results provide a consistent basis for using objective error-reduction methods to strengthen TC track guidance for operational decision-making.

\end{abstract}

\section{Introduction}
Tropical cyclones (TCs) are among the most consequential weather hazards because their strong winds, storm surge, extreme rainfall, and associated flooding can cause substantial loss of life and economic disruption. The forecast track of TCs strongly controls the geographical distribution and timing of these hazards and is therefore central to warning issuance, evacuation planning, resource allocation, and public risk communication \cite{morss2022improving}. In recent years, considerable progress in TC track forecasts has been achieved through upgrades in observing systems, new data assimilation techniques in numerical models, and increased computational capacity \cite{landsea2018have,rappaport2009advances,yamaguchi2017wgne}. These developments have substantially reduced operational track forecast errors and extended useful forecast lead times across multiple ocean basins \cite{heming2019review}. Forecasts now provide more accurate guidance, allowing warning centers to identify threatened regions earlier and support more timely preparedness decisions \cite{rappaport2009advances}.

A fundamental challenge in weather prediction is that the future atmospheric state cannot be described completely by a single deterministic forecast. Uncertainty in the initial conditions, imperfect representations of physical processes, and the nonlinear growth of forecast errors all contribute to a range of plausible atmospheric evolutions. Ensemble prediction systems address this problem by combining multiple forecasts that sample uncertainties in the initial state, model physics, and other forecast-system components, thereby providing flow-dependent information about possible forecast outcomes \cite{leutbecher2008ensemble}. For TC prediction, the ensemble mean can reduce uncorrelated member errors, while ensemble spread can provide information about forecast confidence. Early work showed that averaging dynamical model track forecasts reduced both mean and upper-tail errors relative to individual models \cite{goerss2000tropical}. Later studies of global ensemble systems confirmed the value of ensemble means and probabilistic track guidance \cite{majumdar2010ability,buckingham2010short,yamaguchi2012relative}. However, the practical use of TC ensembles remains challenging. Ensemble
spread is not inherently calibrated, and its relationship with forecast error
varies across basins, lead times, and flow regimes \cite{titley2020global}.
Shared model and data-assimilation frameworks can produce correlated biases,
so small spread does not necessarily imply small track error
\cite{thanh2016application,tsai2013detection}. Computational trade-offs among
ensemble size, resolution, and forecast length limit uncertainty sampling
\cite{leutbecher2008ensemble}, while tracking failures can reduce the effective
ensemble size \cite{buckingham2010short,tsai2013detection}. Distinct track
scenarios and outlying members may also be obscured by, or degrade, the
unweighted ensemble mean \cite{don2016mixture,kowaleski2020use}. Therefore, extracting accurate, reliable, and operationally interpretable guidance from a raw TC ensemble remains a distinct methodological challenge.

In recent years, a range of methods has been developed to extract more reliable TC track
guidance from raw ensembles. These methods can be broadly divided into
statistical and machine-learning (ML) approaches. Statistical methods trim,
select, weight, or correct ensemble members using forecast geometry,
historical performance, recent observations, or trajectory distributions.
They include selective consensus
\cite{elsberry2000consensus,zhang2017probabilistic}, performance-based weighting
and bias correction \cite{vijaya2003multimodel,simon2018description},
short-lead-error selection \cite{qi2014selective,du2016selective},
observation-based subsetting \cite{dong2016obest}, and trajectory clustering
and pruning
\cite{tsai2013detection,don2016mixture,kowaleski2016regression,kowaleski2020use}.
ML methods learn nonlinear relationships between forecast information and
observed track errors. Their predictors may include track position, storm
intensity, pressure, ensemble statistics, and spatial atmospheric fields.
Unlike methods restricted to selecting or weighting existing members, ML
models can generate corrected tracks that are not present in the original
ensemble. Applications include tree-based learning
\cite{tan2021western,boussioux2022hurricane}, neural-network postprocessing
\cite{rasp2018neural,cheung2022hybrid}, and deep architectures for spatial and
temporal forecast information
\cite{cheung2022hybrid,agrawal2025using,giffard2020tropical}. However, reviews of these developments \cite{heming2019review,conroy2023track} indicate that previous methods have generally been evaluated using different forecast systems, basins, periods, lead times, and verification procedures, making their relative accuracy, robustness, and operational applicability difficult to assess directly.

To address this gap, this study presents a controlled comparison of six
objective ensemble uncertainty-reduction approaches spanning member trimming,
trajectory clustering, and ML-based correction. Each method is evaluated relative to the full ensemble mean using 138 TCs during 2016--2024. The forecasts come from three operational global ensemble prediction systems. A common verification framework is applied across 4 initialization times and 25 lead times from 0 to 144 h. Methods requiring historical training
use the same five-fold, case-based cross-validation procedure to produce
out-of-fold predictions.

The analysis addresses three questions: whether objective ensemble correction
consistently improves upon the full ensemble mean, how its effectiveness
varies with forecasting system, lead time, ensemble spread, and individual
case, and whether ML methods provide greater accuracy and robustness than
simpler statistical approaches. Performance is evaluated using track errors, together with case-level improvement frequencies, spread-conditioned performance, an observation-based low-error bound, and the unperturbed control forecast. Sensitivity, data requirements,
and computational considerations are also examined. By holding the forecast
sample and verification procedures constant, this study identifies the
conditions under which each approach improves or degrades TC track guidance
and assesses its potential and limitations for operational use.

\section{Data}

\subsection{Ensemble Forecast Data}

TC ensemble forecasts are obtained from the THORPEX Interactive Grand Global Ensemble (TIGGE) Model Tropical Cyclone Track Data archive \cite{bougeault2010thorpex,ncar_gdex_dataset_d330003}. The analysis includes
three operational global ensemble prediction systems during 2016--2024: the National Centers for Environmental Prediction Global Ensemble Forecast System (GEFS) \cite{zhou2022development}, the UK Met Office Global and Regional Ensemble Prediction System--Global (MOGREPS-G) \cite{bowler2008mogreps}, and the European Centre for Medium-Range Weather Forecasts Integrated Forecasting System (IFS) \cite{palmer2019ecmwf}. The archived track files provide the forecast initialization times, ensemble-member identifiers, storm names, basins, latitudes, longitudes, minimum central pressures, and maximum wind speeds. 

Nominal ensemble size varies among systems and model configurations. IFS
contains 50 perturbed members throughout the study period
\cite{magnusson2019ecmwf}. GEFS contains 20 perturbed members before the
implementation of GEFS v12 on 23 September 2020 and 30 members thereafter
\cite{zhou2022development}. MOGREPS-G contains 24 time-lagged members before
11 July 2017 and 36 members thereafter, following an increase from 12 to 18
members per forecast cycle. The effective ensemble size can be smaller for an
individual case or lead time because members with missing or invalid positions
are excluded.

For each storm and forecast system, the first archived initialization that
exactly matches the cyclone name and occurs within its observed lifetime
defines the 0-h initialization time. Additional initializations 24, 48, and
72 h later are selected when available. Later
initialization times are excluded because the available sample decreases substantially beyond 72 h. Each forecast trajectory is evaluated at 6-h intervals from 0 to 144 h, giving 25 forecast lead times.

\subsection{Observational Track Data}

Observed TC tracks are obtained from the International Best Track Archive for
Climate Stewardship (IBTrACS)
\cite{knapp2010international,gahtan2024ibtracs}. The analysis includes all TCs
during 2016--2024 that attain Category 4 or 5 intensity on the Saffir--Simpson
hurricane wind scale \cite{simpson1974hurricane}. This criterion focuses the
analysis on the most intense storms and provides sufficiently complete
forecast records for comparison among systems. Forecast disturbances are matched to IBTrACS storms by cyclone name, and each
forecast initialization is required to occur within the corresponding
best-track lifetime. The observational tracks are sampled at 6-h intervals to
match the forecast data. Forecast and best-track positions are paired only
when their valid times match exactly, and no temporal interpolation is applied.

A case is defined as a unique combination of forecast system, TC, and
initialization time. The final dataset contains 138 TCs and 1,181 forecast
cases, corresponding to 29,525 case--lead combinations.
Table~\ref{tab:data_summary} summarizes the cases by forecast system and
initialization time. Table~\ref{tab:tc_names_year_basin} lists the names and
basins of all TCs included in the analysis.

\begin{table}[t]
\centering
\caption{Numbers of retained TC forecast cases by ensemble
prediction system and initialization time. Initialization times are
measured relative to the first selected forecast for each storm and system.}
\label{tab:data_summary}
\begin{tabular}{lrrrrr}
\hline
\multirow{2}{*}{Forecast System} & \multicolumn{4}{c}{Initialization Time} & \multirow{2}{*}{Total} \\
\cline{2-5}
 & 0 h& 24 h& 48 h& 72 h& \\
\hline
GEFS        & 134 & 111 & 92 & 73 & 410 \\
MOGREPS-G   & 135 & 118 & 98 & 81 & 432 \\
IFS         & 131 &  99 & 69 & 40 & 339 \\
\hline
All Systems & 400 & 328 & 259 & 194 & 1181 \\
\hline
\end{tabular}
\end{table}

\section{Methods}
\label{sec:methods}
This section defines the full ensemble mean used as the reference TC track
forecast and its verification metrics, followed by the six objective ensemble
uncertainty-reduction approaches evaluated in this study.

\subsection{Full Ensemble Mean}

Let $c\in C$ denote a forecast case, $h\in H=\{0,24,48,72\}$ h the initialization time, $m\in M=\{1,2,3,...\}$ a perturbed ensemble member, and $t\in T=\{0,6,\ldots,144\}$ h the forecast lead time. Unless otherwise specified, verification results are averaged over the four initialization times $h$. The forecast position of member $m$ is denoted by $\hat{\mathbf{x}}_{cmt}=(\widehat{\mathrm{LAT}}_{cmt},\widehat{\mathrm{LON}}_{cmt})^{\mathsf T}$, and the verifying best-track position is $\mathbf{y}_{ct}=(\mathrm{LAT}_{ct},\mathrm{LON}_{ct})^{\mathsf T}$. The full ensemble-mean track used as the reference forecast is the coordinatewise arithmetic mean
\begin{equation}
\overline{\hat{\mathbf{x}}}_{ct}
=\frac{1}{M}\sum_{m=1}^{M}\hat{\mathbf{x}}_{cmt},
\label{eq:full_mean}
\end{equation}

Distances and track errors are calculated using the great-circle distance
\begin{equation}
d_{\mathrm{gc}}(\mathbf{p}_1,\mathbf{p}_2)
=2R\sin^{-1}\!\left[
\sqrt{\sin^2\!\left(\frac{\Delta\mathrm{LAT}}{2}\right)
+\cos(\mathrm{LAT}_1)\cos(\mathrm{LAT}_2)
\sin^2\!\left(\frac{\Delta\mathrm{LON}}{2}\right)}
\right],
\label{eq:haversine}
\end{equation}
where $\mathbf{p}_i=(\mathrm{LAT}_i,\mathrm{LON}_i)^{\mathsf T}$, $\Delta\mathrm{LAT}=\mathrm{LAT}_2-\mathrm{LAT}_1$, $\Delta\mathrm{LON}=\mathrm{LON}_2-\mathrm{LON}_1$, all coordinates are expressed in radians, and $R=6371$ km.

For each case $c$, the full ensemble-mean track error at lead time $t$ is
\begin{equation}
e_{ct}
=d_{\mathrm{gc}}\!\left(
\overline{\hat{\mathbf{x}}}_{ct},
\mathbf{y}_{ct}
\right).
\label{eq:full_mean_error}
\end{equation}
The mean track error across all cases at lead time $t$ is
\begin{equation}
\overline{e}_{t}
=\frac{1}{C}\sum_{c=1}^{C}e_{ct}
=\frac{1}{C}\sum_{c=1}^{C}
d_{\mathrm{gc}}\!\left(
\overline{\hat{\mathbf{x}}}_{ct},
\mathbf{y}_{ct}
\right).
\label{eq:full_mean_case_average}
\end{equation}

\subsection{Dynamic Trimming (DT)}

DT is motivated by evidence that small, distant track branches can represent spurious outliers \cite{tsai2013detection}. The method
implements the selective-consensus principle
\cite{elsberry2000consensus} by identifying geometrically anomalous members,
removing their complete trajectories, and calculating a new ensemble mean
from the retained members. Related approaches include representative-member selection
\cite{molteni2001strategy,weusthoff2011best,liu2024new}, robust trimmed forecast averaging \cite{jose2008simple}, and selective ensemble pruning \cite{caruana2004ensemble,partalas2010ensemble}.

For each lead time $t$, the coordinatewise ensemble median
$\widetilde{\hat{\mathbf{x}}}_{t}$ and the full ensemble mean
$\overline{\hat{\mathbf{x}}}_{t}$ are calculated. Let $t_j=6(j-1)$ h
denote the $j$th forecast lead, where $j=1,\ldots,T$. Each
member receives the score
\begin{equation}
\resizebox{0.9\linewidth}{!}{$\displaystyle
S^{\mathrm{DT}}_{m}
=\frac{1}{3}\left[
\underbrace{
\frac{1}{T}\sum_{j=1}^{T}
d_{\mathrm{gc}}\!\left(
\hat{\mathbf{x}}_{m,t_j},
\widetilde{\hat{\mathbf{x}}}_{t_j}
\right)
}_{\substack{\text{displacement}\\
\text{from the ensemble median}}}
+
\underbrace{
\frac{1}{T}\sum_{j=1}^{T}
d_{\mathrm{gc}}\!\left(
\hat{\mathbf{x}}_{m,t_j},
\overline{\hat{\mathbf{x}}}_{t_j}
\right)
}_{\substack{\text{displacement}\\
\text{from the ensemble mean}}}
+
\underbrace{
\frac{1}{T_{72\text{--}144}}\sum_{j=13}^{T}
d_{\mathrm{gc}}\!\left(
\hat{\mathbf{x}}_{m,t_j},
\widetilde{\hat{\mathbf{x}}}_{t_j}
\right)
}_{\substack{\text{long-range displacement}\\
\text{from the ensemble median}
}}
\right].
$}
\label{eq:dt_score}
\end{equation}

The three terms in Eq.~\eqref{eq:dt_score} quantify each member's displacement from the central tendency of the ensemble. The third term gives additional weight to divergence during the 72--144-h period, when ensemble trajectories tend to separate more strongly. DT is termed \emph{dynamic} because these scores and member rankings are recalculated
from the trajectory geometry of every forecast case rather than being fixed according to member identity or historical performance.

Members are ranked by $S^{\mathrm{DT}}_{m}$, with larger scores indicating
greater displacement from the ensemble family. DT uses hard trimming rather
than continuous score-based weighting because preliminary experiments show
larger error reductions from trimming. The sensitivity analysis in
Figure~\ref{Fig:sensitivity_dt_removal_sensitivity_medium_long} shows that
performance improves as the trimming fraction increases to 20\% and provides
no further improvement at larger fractions. The same trimming fraction is used for
the other trimming methods to provide a consistent comparison. Because the
ranking uses only the geometry of the current forecast trajectories, it is directly applicable in real-time TC ensemble forecasting.

\subsection{Historical-performance Trimming (HIST)}

HIST determines member contributions from performance in historical
forecast--verification cases. The method assumes that the relative skill
associated with an ensemble member has some temporal persistence:
a member that consistently produces large historical track errors is less
likely to provide useful guidance in a new forecast and is therefore excluded
before calculating the corrected ensemble mean. The usefulness of HIST depends on how the ensemble is constructed. In an
idealized ensemble whose members differ only through random, exchangeable
initial-condition perturbations, a fixed historical ranking of individual
members would have little physical basis. HIST is more relevant when member
identity may reflect repeatable differences arising from the design of the
forecast system, such as stochastic or perturbed physics, cycling strategy, or
other sources of model uncertainty. Forecast-system upgrades may also affect HIST by changing model physics,
resolution, data assimilation, perturbation methods, or member configuration,
which can weaken the temporal consistency of member-specific historical skill.
Accordingly, the historical ranking used here should be interpreted as
system-dependent and most applicable within periods over which the ensemble
configuration is relatively stable. This approach is related to historical member-skill assessment and exclusion \cite{doycheva2017assessment}, training-period
superensemble weighting \cite{vijaya2003multimodel,krishnamurti2000multimodel}, and previous-season statistical weighting \cite{weber2003hurricane}.

We apply a five-fold cross-validation procedure to prevent the predicted cases from entering the member ranking. For each held-out
fold, the mean great-circle track error of each member is first calculated in every case belonging to the other four
folds. These case-level errors are then averaged separately by ensemble member to obtain the fold-specific historical-performance score $S^{\mathrm{HIST}}_{m}$. A smaller score indicates greater historical track accuracy. For each held-out case, the available members are ranked according to their
historical-performance scores. The 20\% with the largest scores are removed,
consistent with the trimming fraction selected by DT. The procedure cycles through all five folds so that every case receives an out-of-fold prediction. Therefore, HIST uses only historical forecast performance and requires neither observations from the current case nor the geometry of its forecast trajectories.

\subsection{Short-lead-error Trimming (SLET)}

SLET adapts short-lead selective ensemble averaging
\cite{qi2014selective} and observation-based ensemble optimization
\cite{dong2016obest,li2020study}. Whereas DT emphasizes long-range divergence,
SLET selects members entirely from their observed track errors during the
early forecast period. It therefore provides a delayed update after the
initial evolution of the storm becomes known.
Once the verifying best-track position at the selection lead $t_s$ became available, each member receives the score
\begin{equation}
S^{\mathrm{SLET}}_{m}
=d_{\mathrm{gc}}\!\left(
\hat{\mathbf{x}}_{m,t_s},
\mathbf{y}_{t_s}
\right),
\qquad t_s=24~\mathrm{h}.
\label{eq:slet_score}
\end{equation}
The selection lead $t_s=24$ h is chosen pragmatically rather than through
parameter optimization. It provides enough time for differences in the early
evolution of the member tracks to become apparent while retaining forecasts
through 144 h. For storms that remain a threat after the selection time, the
updated guidance can therefore still support risk-planning and preparedness
decisions \cite{morss2022improving}. For each case, members are ranked from smallest to largest score. The 20\% with the largest 24-h errors are removed, consistent with DT. To prevent the 24-h position used for member selection from also contributing to verification, SLET is evaluated only for $t>24$ h, corresponding to lead times from 30 to 144 h. Because the required verifying
position is unavailable at the original initialization time, SLET represents
observation-conditioned delayed guidance rather than a forecast available at
initial issuance.

\subsection{Regression-mixture Clustering (RC)}

RC combines regression-mixture trajectory clustering
\cite{gaffney1999trajectory,gaffney2007probabilistic} with objective
small-cluster pruning \cite{kowaleski2020use}. Regression-mixture clustering
has been applied to historical TC tracks \cite{camargo2007cluster},
alternative track scenarios in ensemble forecasts \cite{tsai2013detection},
and single- and multimodel ensemble forecasts
\cite{don2016mixture,kowaleski2016regression}. Applications to regional
ensembles have further demonstrated its ability to relate differences in
forecast tracks to cyclone structural evolution
\cite{gonzalez2018use,kowaleski2018relationship}. Related approaches include
density-based trajectory clustering \cite{oettli2024objective}, Bayesian
forecast-scenario selection \cite{yamaguchi2026optimal}, and recombination
clustering for constructing probabilistic forecasts \cite{liu2023recombination}.

For each matched TC and initialization time, member trajectories from all
available forecast systems are pooled, with each system--member pair treated
as a distinct trajectory. RC
represents each cluster using polynomial latitude and longitude trajectories
and an associated covariance structure. The expectation--maximization (EM)
algorithm iteratively estimates the cluster weights, polynomial coefficients,
covariance matrices, and posterior membership probabilities. Each member is
assigned to the cluster for which it has the largest posterior probability.
Mathematical details of RC are provided by 
\cite{gaffney1999trajectory,gaffney2007probabilistic,don2016mixture}.

After cluster assignment, clusters containing less than 20\% of the pooled
members are removed, adapting the small-cluster pruning approach
\cite{kowaleski2020use}. If this criterion removes every cluster, the largest
cluster is retained. The RC forecast is the equal-weight mean of all members
in the retained clusters at each lead time. Section~\ref{appendix_b} presents
the hyperparameter sensitivity analysis. Cluster fitting uses only the
current forecast trajectories and does not use best-track positions.
Therefore, once its configuration is fixed, RC can be generated at initial
forecast issuance.

\subsection{Convolutional Neural Network (CNN)}

CNNs use shared, learnable filters to extract features from structured inputs
efficiently \cite{lecun1998gradient}. Although originally developed for image
processing, convolution can also operate along a temporal sequence and produce
a corrected track outside the original ensemble. In TC forecasting, CNNs
process gridded atmospheric data for track prediction
\cite{giffard2020tropical,cheung2022hybrid} and satellite observations for
intensity estimation \cite{chen2019estimating,wimmers2019using}.

This study adapts the architecture of 
\cite{agrawal2025using} to forecast tracks through 144 h. The network contains
two one-dimensional temporal convolution layers followed by a Multi-Layer Perceptron (MLP) with three hidden layers and a fully connected output layer. Rectified Linear Unit (ReLU) activation follows each hidden layer. The predictors consist only of the latitude--longitude
sequences of the ensemble members and ensemble mean. For case $c$, the input
is
\begin{equation}
\hat{\mathbf{X}}_{c}
\in\mathbb{R}^{2(M+1)\times T},
\label{eq:cnn_input}
\end{equation}
where the first $2M$ rows contain the member tracks and the final two rows
contain the ensemble-mean track.

Let $f_{\boldsymbol{\theta}}$ denote the complete CNN mapping. Its output is
\begin{equation}
\hat{\mathbf{Y}}_{c}^{\mathrm{CNN}}
=
f_{\boldsymbol{\theta}}\!\left(\hat{\mathbf{X}}_{c}\right)
=
\begin{pmatrix}
\widehat{\mathrm{LAT}}^{\mathrm{CNN}}_{ct_1} & \cdots &
\widehat{\mathrm{LAT}}^{\mathrm{CNN}}_{ct_T} \\
\widehat{\mathrm{LON}}^{\mathrm{CNN}}_{ct_1} & \cdots &
\widehat{\mathrm{LON}}^{\mathrm{CNN}}_{ct_T}
\end{pmatrix},
\label{eq:cnn_output}
\end{equation}
and the predicted position at lead time $t$ is
\begin{equation}
\hat{\mathbf{x}}^{\mathrm{CNN}}_{ct}
=
\left(
\widehat{\mathrm{LAT}}^{\mathrm{CNN}}_{ct},
\widehat{\mathrm{LON}}^{\mathrm{CNN}}_{ct}
\right)^{\mathsf T}.
\label{eq:cnn_position}
\end{equation}

CNN uses the same five-fold partition as HIST, and each input channel is
standardized using training-fold statistics only. Because the original study
does not provide a complete hyperparameter search space, the search used here
is defined in Section~\ref{appendix_b}. Best-track positions are used only as
training targets and are not included among the CNN predictors.

\subsection{Gradient Boosted Trees (GBT)}

GBT combines a sequence of regression trees into a single prediction
\cite{friedman2001greedy}. Each new tree corrects errors remaining from the
preceding trees, allowing the model to represent nonlinear relationships and
interactions among predictors. GBT has previously been applied to TC track
prediction and correction \cite{tan2021western,boussioux2022hurricane}.

Rather than removing ensemble members, GBT predicts eastward and northward
corrections to the full ensemble mean. The transformation $G$ converts the
displacement between two geographical positions into these directional
components:
\begin{equation}
G(\mathbf{p}_1,\mathbf{p}_2)
=
d_{\mathrm{gc}}(\mathbf{p}_1,\mathbf{p}_2)
\begin{pmatrix}
\sin\beta(\mathbf{p}_1,\mathbf{p}_2)\\
\cos\beta(\mathbf{p}_1,\mathbf{p}_2)
\end{pmatrix},
\label{eq:gbt_geographical_transform}
\end{equation}
where $\beta(\mathbf{p}_1,\mathbf{p}_2)$ is the direction from
$\mathbf{p}_1$ to $\mathbf{p}_2$. The inverse transformation $G^{-1}$ moves a
starting position by specified eastward and northward distances. The training
targets are
\begin{equation}
\begin{pmatrix}
r^{E}_{ct}\\
r^{N}_{ct}
\end{pmatrix}
=
G\!\left(
\overline{\hat{\mathbf{x}}}_{ct},
\mathbf{y}_{ct}
\right),
\label{eq:gbt_target}
\end{equation}
where $r^{E}_{ct}$ and $r^{N}_{ct}$ denote the required eastward and
northward corrections.

Let $\mathbf{z}_{ct}$ denote the forecast predictor vector listed in
Table~\ref{tab:ml_new_track_predictors}. The predictors summarize forecast
timing, ensemble size, position and spread, track motion and displacement,
initialization date, and storm intensity. For each direction
$q\in\{E,N\}$, the predicted correction is
\begin{equation}
f_{\boldsymbol{\theta}_q}(\mathbf{z}_{ct})
=
F_q^{(0)}
+
\eta\sum_{b=1}^{B}h_{qb}(\mathbf{z}_{ct}),
\label{eq:gbt_model}
\end{equation}
where $F_q^{(0)}$ is the initial prediction, $h_{qb}$ is the correction from
the $b$th tree, $\eta$ is the learning rate, and $B$ is the number of boosting
iterations. The corrected position is therefore
\begin{equation}
\hat{\mathbf{x}}^{\mathrm{GBT}}_{ct}
=
G^{-1}\!\left[
\overline{\hat{\mathbf{x}}}_{ct},
F_E^{(0)}+\eta\sum_{b=1}^{B}h_{Eb}(\mathbf{z}_{ct}),
F_N^{(0)}+\eta\sum_{b=1}^{B}h_{Nb}(\mathbf{z}_{ct})
\right].
\label{eq:gbt_track}
\end{equation}

GBT uses the same five-fold case partition as HIST. Missing predictor values
are replaced by training-sample medians, and separate models are fitted for
each forecasting system, initialization time, and correction direction. Section~\ref{appendix_b} describes the training and hyperparameter search. Best-track positions define the training targets but are not predictors, so the trained models require only ensemble forecast information at inference.

\begin{figure}[htbp]
    \centering
    \includegraphics[scale=0.4]{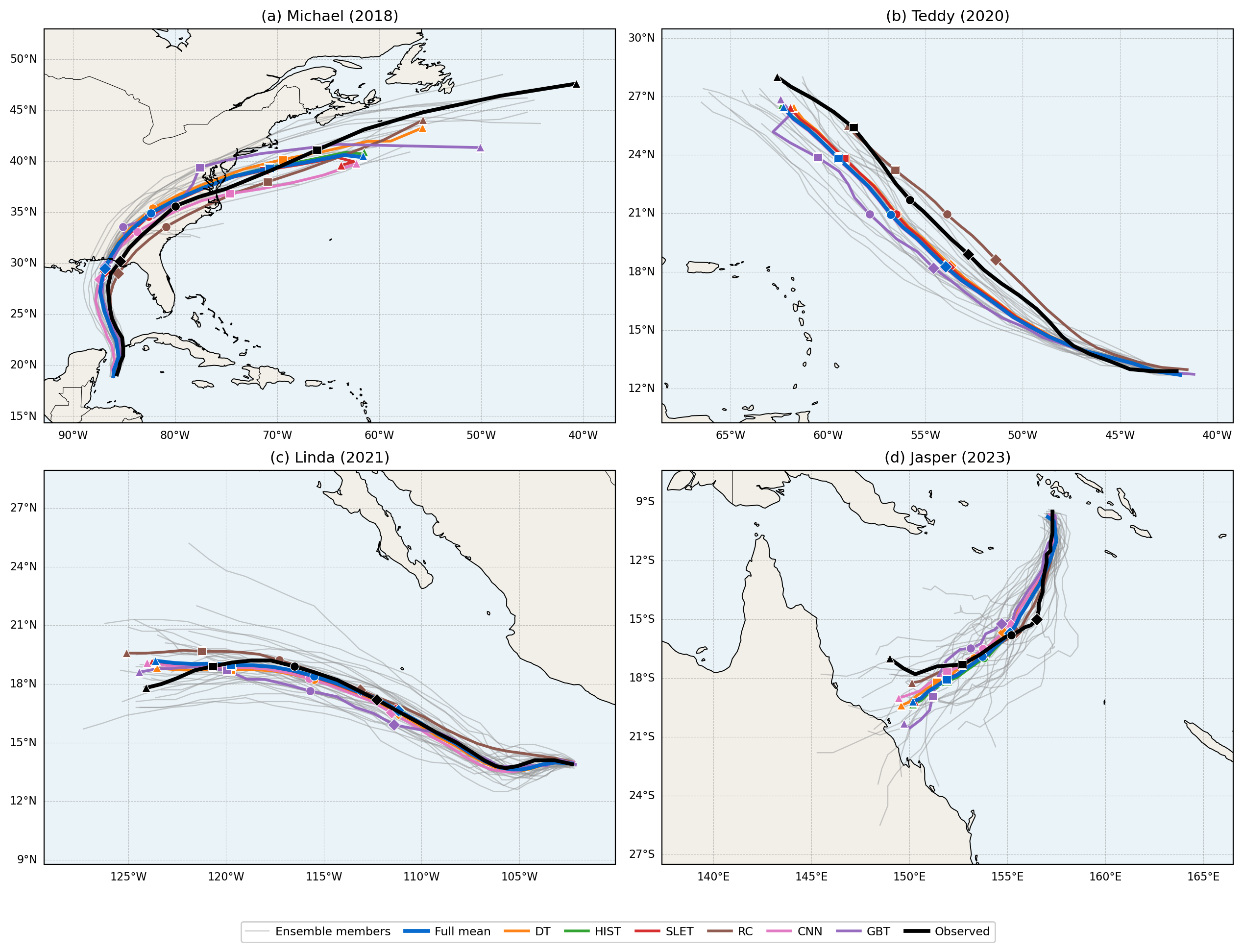}
    \caption{Illustrative TC track forecasts from GEFS at the 0-h initialization time. Light-gray lines show individual ensemble members, while the full ensemble mean, corrected forecasts, and observed track are identified in the legend. Diamond, circle, square, and triangle markers indicate forecast positions at 72, 96, 120, and 144 h, respectively.}
    \label{Fig:case_study_four_tcs_tracks_ml_min_leaf_1}
\end{figure}

Figure~\ref{Fig:case_study_four_tcs_tracks_ml_min_leaf_1} illustrates the ensemble-track inputs and the resulting corrected tracks for four representative cases. The observed track is included as the verification reference, while the individual members and full ensemble mean provide the common inputs and baseline for the correction methods.

\section{Results}

\subsection{Baseline Ensemble Error and Spread}

\begin{figure}[htbp]
    \centering
    \includegraphics[scale=0.4]
    {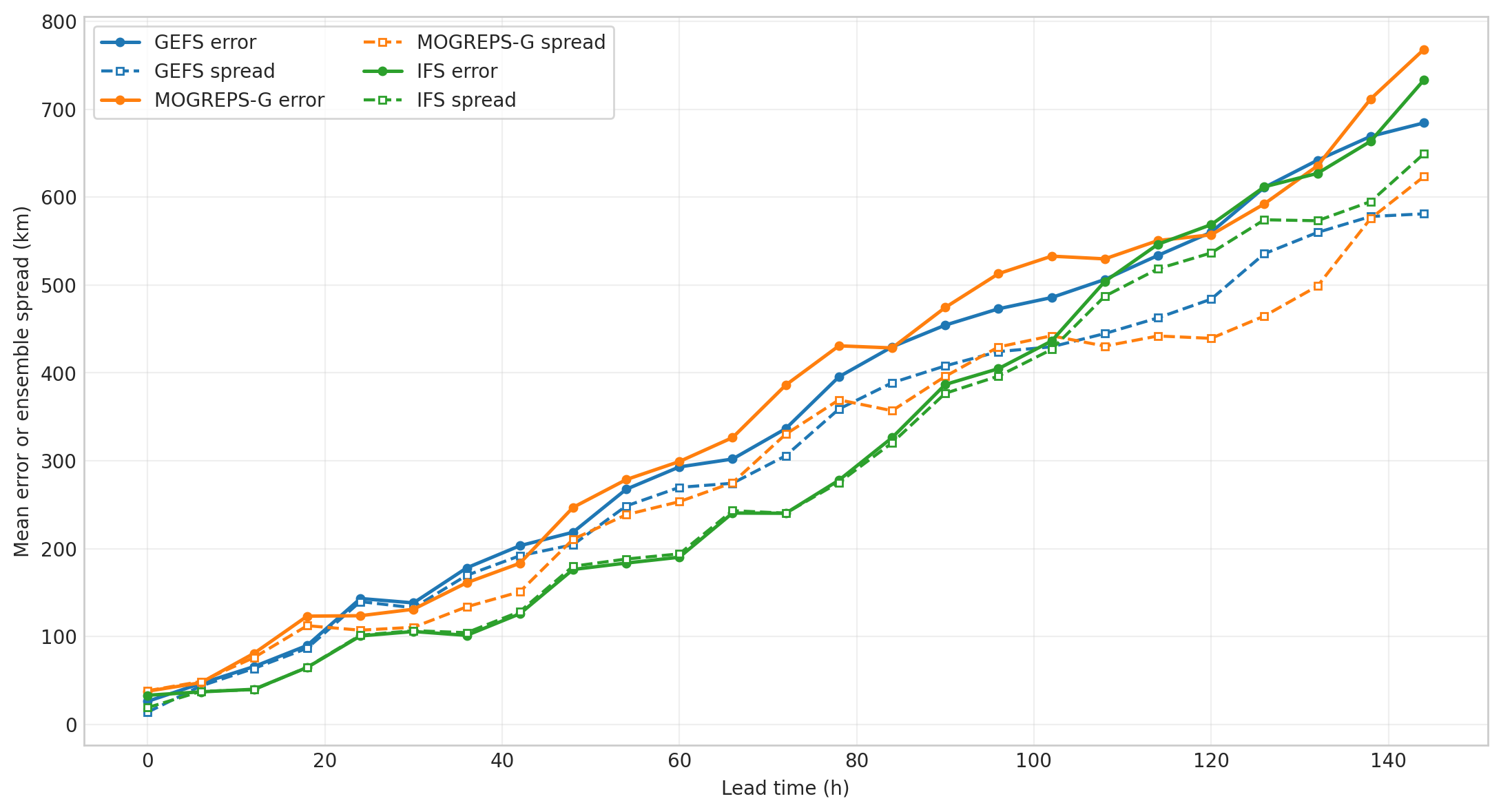}
    \caption{Mean full-ensemble track error and radial ensemble spread as
functions of forecast lead time for GEFS, MOGREPS-G, and IFS. Track error
(solid lines) is the great-circle distance between the ensemble-mean and
best-track positions. Ensemble spread (dashed lines) is the case-level mean
great-circle distance between the perturbed members and ensemble mean.}
    \label{Fig:full_mean_error_by_center_and_lead}
\end{figure}

Before comparing the uncertainty-reduction methods, we examine the baseline
track errors of the full ensemble mean and the ensemble spread of the three
forecasting systems. As shown in
Figure~\ref{Fig:full_mean_error_by_center_and_lead}, the full ensemble-mean
track error increases substantially with forecast lead time for all three
systems. Mean errors at 0 h are 26.4 km for GEFS, 37.9 km for MOGREPS-G, and
33.2 km for IFS. By 144 h, these errors reach 684.4, 768.0, and
733.1 km, respectively. The case-mean spread distributions have broadly similar medians among systems
(Figure~\ref{Fig:ensemble_spread_boxplot_by_center}). Spread increases with
lead time but generally grows more slowly than ensemble-mean error,
particularly at longer leads. Spread remains relatively close to error for
IFS through much of the short and medium range, whereas larger differences
develop for GEFS and MOGREPS-G. This behavior is consistent with increasing
long-range underdispersion. Previous studies identify long-range
underdispersion in ECMWF ensembles \cite{majumdar2010ability},
basin-dependent underdispersion in GEFS \cite{buckingham2010short}, and
insufficient along-track dispersion in several operational systems
\cite{leonardo2017verification}. Possible causes include incomplete sampling
of initial-condition uncertainty \cite{majumdar2010ability}, systematic errors
shared among members \cite{leonardo2017verification}, and incomplete
representation of model uncertainty \cite{zhou2022development}.

\subsection{Overall Performance across Methods}
\label{sec:Overall Performance across Methods}

\begin{table}[htbp]
\centering
\caption{Improvement in TTE, ATE, and
CTE relative to the full ensemble mean for short
(0--48 h), medium (54--96 h), and long (102--144 h) forecast ranges. Entries show
the percentage reduction in mean error, with the absolute reduction in km in
parentheses; negative values indicate degradation.}
\label{tab:tte_ate_cte_improvement_by_range}
\begin{tabular}{lrrr}
\toprule
Method & TTE improvement & ATE improvement & CTE improvement \\
\midrule
\multicolumn{4}{c}{Short range (0--48 h)} \\
\midrule
DT   &  8.0\% (9.1)    &  7.7\% (6.3)  &  9.6\% (7.3)  \\
HIST & -0.3\% (-0.3)  &  0.0\% (0.0)  & -0.3\% (-0.2) \\
SLET & 11.3\% (18.9)  & 10.3\% (11.3) & 14.5\% (14.9) \\
RC   &  2.5\% (2.9)    &  0.4\% (0.3)  &  6.2\% (4.7)  \\
CNN  &  0.2\% (0.2)    &  0.1\% (0.1)  &  1.0\% (0.8)  \\
GBT  & -12.8\% (-14.6) &  1.8\% (1.4)  & -4.2\% (-3.2) \\
\midrule
\multicolumn{4}{c}{Medium range (54--96 h)} \\
\midrule
DT   & 13.4\% (47.1) & 11.8\% (28.0) & 15.4\% (32.2) \\
HIST &  1.4\% (5.1)  & -0.2\% (-0.6) &  2.0\% (4.1)  \\
SLET &  5.5\% (19.3) &  6.1\% (14.5) &  5.5\% (11.6) \\
RC   &  2.1\% (7.4)  &  3.4\% (8.1)  & -1.2\% (-2.5) \\
CNN  &  3.7\% (12.9) &  5.5\% (13.2) &  2.3\% (4.9)  \\
GBT  & 23.4\% (82.5) & 25.9\% (61.2) & 21.8\% (45.6) \\
\midrule
\multicolumn{4}{c}{Long range (102--144 h)} \\
\midrule
DT   & 16.1\% (95.8) & 14.9\% (61.7) & 17.9\% (60.2) \\
HIST &  1.5\% (8.8)  &  0.6\% (2.4)  &  2.3\% (7.7)  \\
SLET &  2.4\% (14.4) &  1.7\% (6.9)  &  3.6\% (12.0) \\
RC   &  2.3\% (13.9) &  1.9\% (7.8)  &  2.6\% (8.8)  \\
CNN  &  1.9\% (11.4) &  0.9\% (3.8)  &  2.9\% (9.9)  \\
GBT  &  8.8\% (52.3) &  8.4\% (34.7) &  9.3\% (31.3) \\
\bottomrule
\end{tabular}
\end{table}

In this section, the overall performance of the six uncertainty-reduction methods is evaluated relative to the full ensemble mean. 
Table~\ref{tab:tte_ate_cte_improvement_by_range} compares the six methods by
forecast range. DT reduces total track error (TTE) by 8.0\% in the short range, 13.4\% in the
medium range, and 16.1\% in the long range, showing that its benefit increases
with lead time. GBT degrades short-range TTE by 12.8\%, produces the largest
medium-range improvement of 23.4\%, and retains an 8.8\% improvement in the
long range. SLET reduces TTE by 11.3\% over its available short-range leads
(30--48 h), but its improvement decreases to 5.5\% and 2.4\% in the medium
and long ranges. HIST, RC, and CNN remain close to the full ensemble mean,
with TTE improvements no greater than 3.7\%.

DT consistently reduces both directional error components, with larger cross-track error (CTE)
than along-track error (ATE) improvements in every forecast range. ATE and CTE are defined as the absolute projections parallel and perpendicular to the observed 6-h motion direction of TTE. This result indicates that DT is particularly effective at reducing displacement perpendicular to the
observed track. GBT produces the largest medium-range reductions in both ATE and CTE, but its short-range degradation limits its consistency. Thus, DT provides the most stable improvement across forecast ranges, whereas GBT
produces larger but more lead-dependent changes.

The pooled lead-specific errors in
Figure~\ref{Fig:method_error_by_lead_all_centers} show the same behavior. DT
generally remains below the full ensemble mean, and its advantage increases
with lead time. GBT produces large errors at the earliest leads, performs best
over much of the medium range, and loses part of its improvement toward
144 h. SLET begins at 30 h, while HIST, RC, and CNN generally remain close to
the full ensemble mean. The center-specific tables further show that RC, CNN,
and GBT vary more strongly among forecasting systems, whereas DT provides the
most consistent long-range improvement.

\subsection{Robustness across Forecast Systems and Cases}

\begin{figure}[htbp]
    \centering
    \includegraphics[scale=0.4]{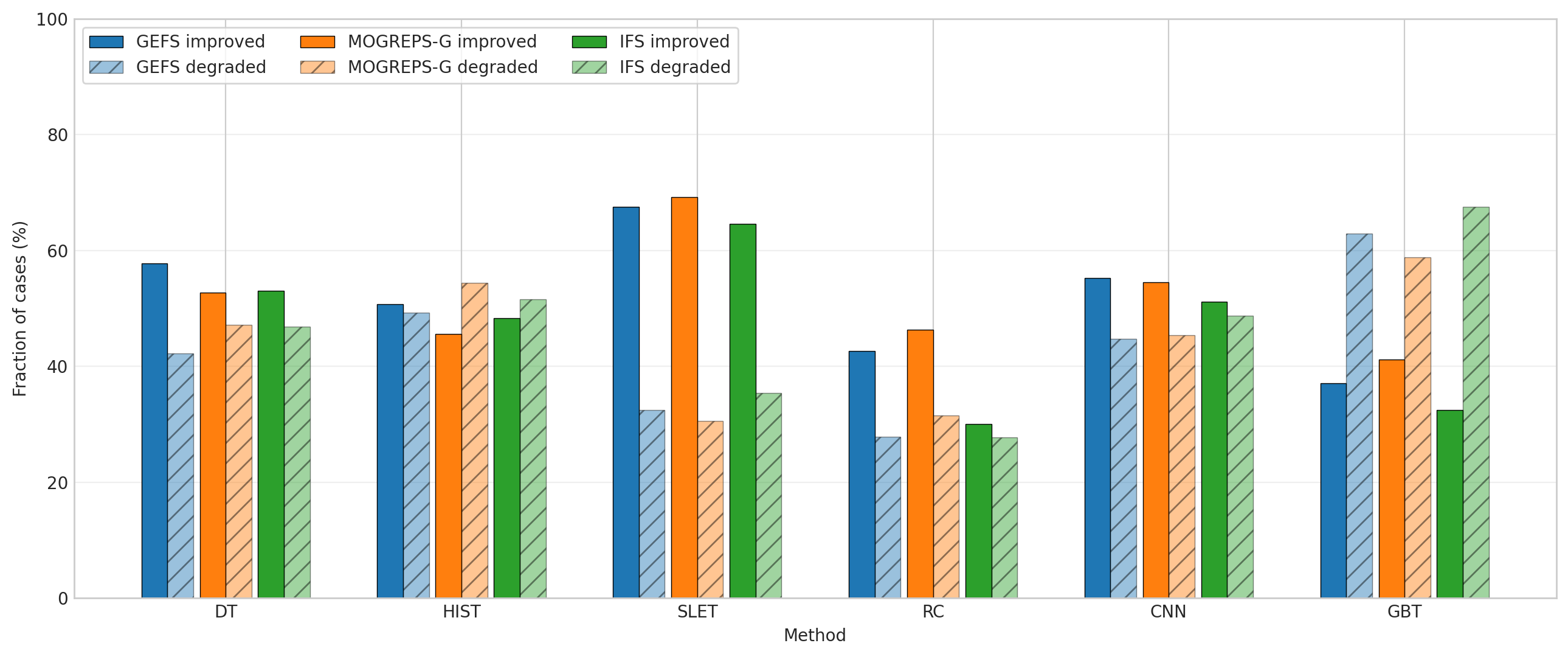}
\caption{Fractions of TC--initialization cases improved (solid bars) or degraded (hatched bars) relative
to the full ensemble mean for GEFS, MOGREPS-G, and IFS. SLET is evaluated only from 30 to 144 h because it uses the observed 24-h position for member selection. Fractions that do not sum to 100\% represent unchanged cases. Cases are pooled across the four initialization
times.}
\label{Fig:case_improved_degraded_fraction_three_centers}
\end{figure}

\begin{table}[htbp]
\centering
\caption{Average track-error improvement relative to the full ensemble mean across GEFS, MOGREPS-G, and IFS for low-, medium-, and high-spread cases. Spread categories are equal-frequency terciles defined separately within each forecasting model and lead time.}
\label{tab:spread_dependence_low_medium_high}

\begin{tabular}{lrrrrrr}
\toprule
Spread category & DT & HIST & SLET & RC & CNN & GBT \\
\midrule
Low spread    & -0.3\% & -0.1\% & 1.9\% & 5.6\% & 1.3\% & -21.9\% \\
Medium spread &  0.0\% & -0.2\% & 2.0\% & 2.4\% & 0.6\% & -17.7\% \\
High spread   & 21.5\% &  1.9\% & 5.0\% & 0.5\% & 2.7\% & 25.0\% \\
\bottomrule
\end{tabular}%
\end{table}

In this section, we evaluate the robustness of each uncertainty-reduction method. The analysis examines whether improvements are consistent among forecasting systems and quantifies the fractions of cases that improve or degrade relative to the full ensemble mean. Figure~\ref{Fig:case_improved_degraded_fraction_three_centers} compares the
fractions of individual cases that improve or degrade relative to the full ensemble mean. SLET improves the largest fraction of
eligible cases across all three forecasting systems, ranging from 64.6\% to
69.2\%. DT also shows consistent case-level performance, improving
52.8--57.8\% of cases and improving more cases than it degrades for every
system. RC leaves many cases unchanged because no members are removed when a
qualifying small cluster is absent. In contrast, GBT degrades 58.8--67.6\%
of cases. Its large medium-range mean
improvement in Table~\ref{tab:tte_ate_cte_improvement_by_range} therefore results from substantial gains in a smaller subset of cases rather than
uniform improvement.

In Table \ref{tab:spread_dependence_low_medium_high}, we show that forecast improvement depends strongly on ensemble spread. For high-spread cases, GBT and DT produced the largest error reductions, improving the full ensemble mean by 25.0\% and 21.5\%, respectively. Large spread provides greater opportunity for correction because divergent members or competing track scenarios can displace the ensemble mean; conversely, trimming tightly clustered members produces little change. This behavior is consistent with \cite{elsberry2000consensus}, who found that selective consensus offers limited benefit for small-spread forecasts but greater potential in large-spread cases.However, GBT degrades low- and medium-spread forecasts by 21.9\% and 17.7\%, indicating overly large
corrections when ensemble uncertainty is limited. RC performs best in the
low-spread category but provides little high-spread improvement, possibly
because the observed trajectory can belong to a minority cluster during track
bifurcation. DT therefore provides the more consistent spread-dependent
behavior, while GBT combines larger high-spread gains with greater degradation
at lower spread.

The lead-specific results for GEFS, MOGREPS-G, and IFS are provided in Tables~\ref{tab:gefs_six_method_improvement},
\ref{tab:mogreps_g_six_method_improvement}, and
\ref{tab:ifs_six_method_improvement}, respectively. DT improves nearly every
lead after 24 h and reduces long-range errors by approximately 12--21\%
across the three systems. SLET improves most eligible leads, although its
benefit generally decreases toward 144 h and becomes slightly negative for
IFS at the final leads. GBT produces large medium-range improvements for GEFS
and MOGREPS-G but is less consistent for IFS and outside the medium range.
HIST remains close to the full ensemble mean, while RC and CNN vary more
strongly among systems.

\subsection{Performance Relative to the Low-error Bound}

Percentage improvement alone does not indicate how much of the potentially correctable ensemble error is captured, because the available improvement differs among forecasting systems and lead times. We therefore compare each method against a low-error bound derived from observational tracks. Although this bound is not available operationally, it provides a diagnostic reference for evaluating how closely each real-time-applicable method approaches the best performance obtainable under the prescribed member-selection framework. The analysis focuses on 72--144 h, when track uncertainty is greatest and objective correction is most valuable.

For each case $c$ and lead time $t$, the number of trimmed members is
$r=0.2M$, consistent with the fraction used by the trimming methods. For a
candidate removal set $\mathcal{R}\subset\{1,\ldots,M\}$, the retained-member
mean is
\begin{equation}
\overline{\hat{\mathbf{x}}}_{ct}(\mathcal{R})
=
\frac{1}{M-r}
\sum_{\substack{m=1\\m\notin\mathcal{R}}}^{M}
\hat{\mathbf{x}}_{cmt},
\end{equation}
where $\hat{\mathbf{x}}_{cmt}$ is the forecast position of member $m$. Using
the verifying best-track position $\mathbf{y}_{ct}$, the optimal removal set is
\begin{equation}
\mathcal{R}^{*}_{ct}
=
\underset{
\substack{
\mathcal{R}\subset\{1,\ldots,M\}\\
|\mathcal{R}|=r
}
}{\arg\min}
\;
d_{\mathrm{gc}}\!\left[
\overline{\hat{\mathbf{x}}}_{ct}(\mathcal{R}),
\mathbf{y}_{ct}
\right].
\end{equation}
The corresponding low-error-bound error is
\begin{equation}
E_{\mathrm{LEB}}(t)
=
\frac{1}{|C_t|}
\sum_{c\in C_t}
d_{\mathrm{gc}}\!\left[
\overline{\hat{\mathbf{x}}}_{ct}
\!\left(\mathcal{R}^{*}_{ct}\right),
\mathbf{y}_{ct}
\right],
\end{equation}
where $C_t$ is the set of cases available at lead time $t$. Because this
optimization is repeated independently at every lead time, the identities of
the removed members may change along the forecast track.

The percentage of the low-error-bound error reduction achieved by method $a$
is
\begin{equation}
LEB_a(t)
=
100\,
\frac{E_{\mathrm{full}}(t)-E_a(t)}
     {E_{\mathrm{full}}(t)-E_{\mathrm{LEB}}(t)}.
\end{equation}
A value of 0\% indicates no improvement over the full ensemble mean, whereas
100\% indicates performance equal to the observation-based 20\% trimming
bound. Because the low-error bound is based on member trimming, the ML methods
can exceed 100\% because they generate new forecast tracks and are not
restricted to selecting a subset of the original members.

\begin{figure}[htbp]
    \centering
    \includegraphics[scale=0.3]{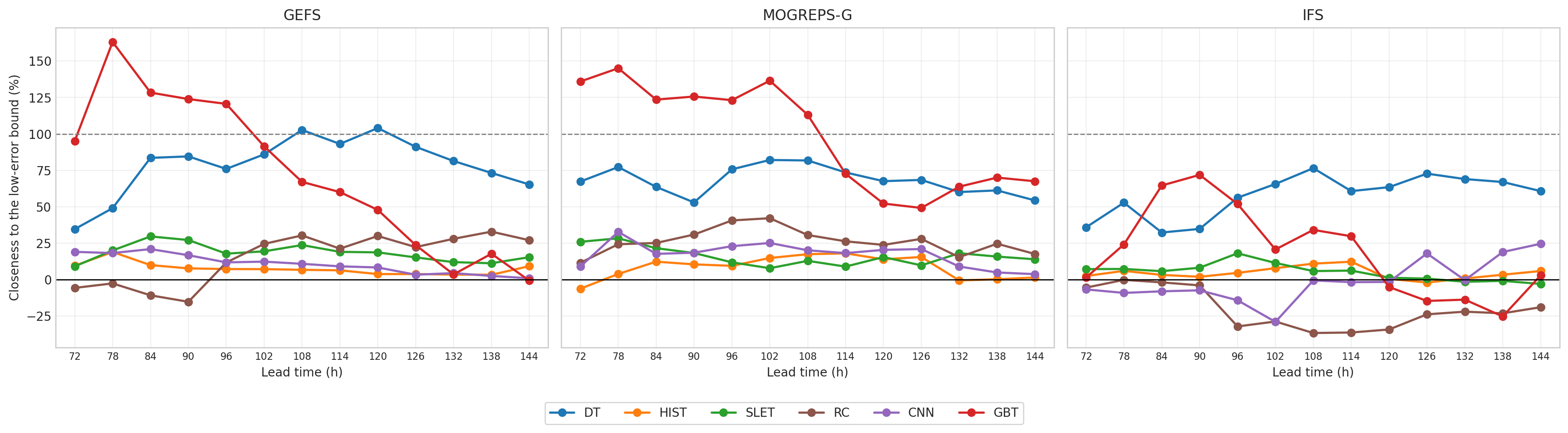}
    \caption{Percentage of the observation-based, lead-specific 20\% trimming gap closed by each method at 72--144-h lead times for GEFS, MOGREPS-G, and IFS. Values of 0\% and 100\% indicate performance equal to the full ensemble mean and the low-error bound, respectively; negative values indicate degradation.}
    \label{fig:oracle_gap_72_144h_three_centers_by_lead}
\end{figure}

As shown in
Figure~\ref{fig:oracle_gap_72_144h_three_centers_by_lead}, when aggregated
over all matched case--lead pairs from 72 to 144 h, DT achieves 79.2\% of the
low-error-bound error reduction for GEFS, 67.3\% for MOGREPS-G, and 60.2\%
for IFS. This indicates that DT provides the most consistent and stable approximation to the low-error bound across the three forecasting systems. GBT achieves 60.9\% of the low-error-bound error reduction for GEFS and 93.0\% for MOGREPS-G, but only 13.0\% for IFS. At certain lead times, GBT can exceed the low-error bound, indicating its potential in providing a more accurate track correction than the trimming methods. Its lead-specific percentages decrease after the maxima, which is consistent with the result in Section \ref{sec:Overall Performance across Methods}. Thus, GBT sometimes exceeds the low-error-bound error reduction but shows substantially weaker consistency across forecasting
systems and lead times.

The remaining methods generally achieve less than 30\% of the error reduction
represented by the low-error bound and occasionally produce larger errors than
the full ensemble mean. Overall, DT performs most consistently relative to the
low-error bound, whereas GBT attains higher percentages in some forecasts but
shows substantially greater variability.

\subsection{Performance Relative to the Control Forecast}

\begin{figure*}[htbp]
    \centering
    \includegraphics[width=\textwidth]{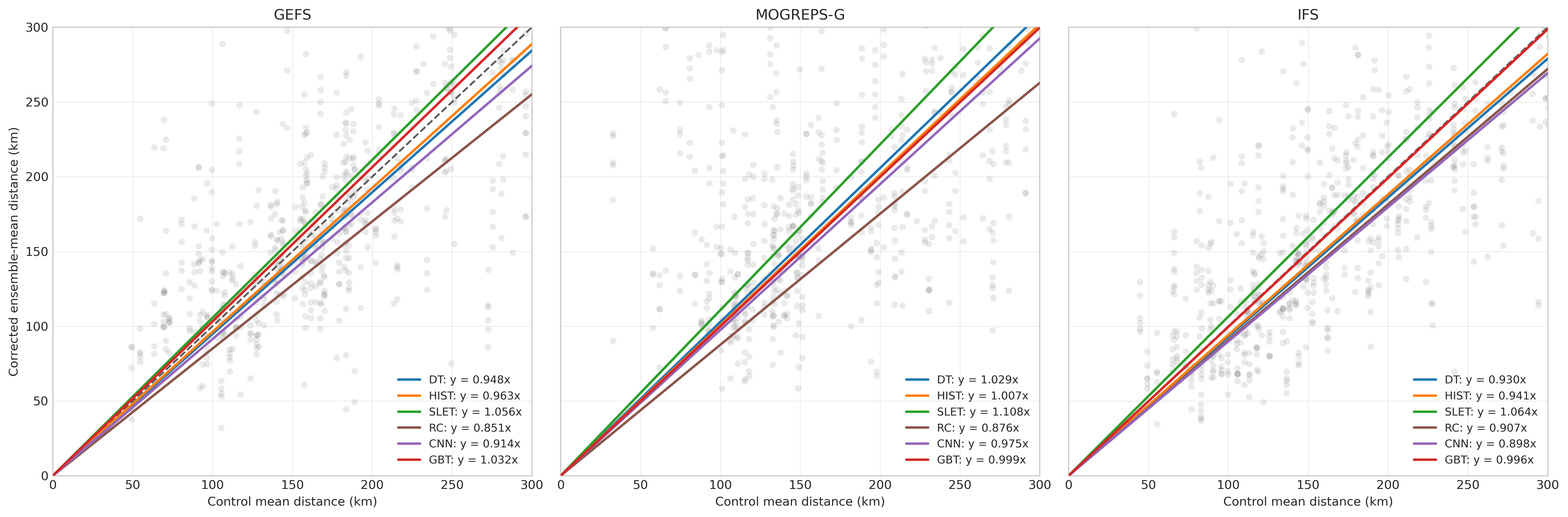}
    \caption{Corrected ensemble-mean track error versus control-forecast track error for GEFS, MOGREPS-G and IFS. Each gray point represents one matched TC forecast case, with errors averaged over its available lead times. The dashed line denotes equal error, and the colored lines are least-squares fits constrained through the origin. Slopes below one indicate that the corrected ensemble mean generally has lower error than the control forecast.}
    \label{fig:corrected_vs_control}
\end{figure*}

As a final evaluation, the corrected ensemble forecasts are compared with the corresponding unperturbed control forecasts. This analysis determines whether improvements relative to the full ensemble mean also translate into performance competitive with an operationally available deterministic forecast. In Figure~\ref{fig:corrected_vs_control}, points below the 1:1 line indicate cases in which the corrected ensemble mean has a lower mean track error than the control forecast.

RC provides the most consistent improvement relative to the control forecast, with fitted slopes of 0.851, 0.876, and 0.907 for GEFS, MOGREPS-G, and IFS, respectively. CNN also produces slopes below one for all three systems (0.914, 0.975, and 0.898), indicating broadly consistent improvements. DT and HIST outperform the control for GEFS and IFS but are slightly worse for MOGREPS-G, where their slopes are 1.029 and 1.007. GBT is approximately neutral relative to the control, with slopes close to one for all three systems. The substantial scatter around the fitted lines nevertheless shows that performance remains case-dependent. Overall, RC and CNN provide the strongest cross-system performance relative to the control forecast, complementing the earlier comparisons with the full ensemble mean.

\section{Discussion}
\subsection{Implications for Operational Forecasting}
This study compares six approaches for reducing ensemble uncertainty in TC
track forecasts. These methods differ in their data requirements and
operational roles, ranging from simple statistical trimming to ML-based track
prediction. Operational TC forecasting combines multiple forms of numerical
guidance with forecaster judgment \cite{rappaport2009advances,conroy2023track},
and forecast products must support a range of decision timelines
\cite{morss2022improving}. The results therefore support role-specific use of
these methods rather than replacing the full ensemble with a single corrected
track. Their utility depends on lead time, ensemble spread, and forecasting system.

DT is the strongest candidate for use at initial forecast issuance. Its error
reduction increases with forecast lead time and remains consistent across the
three forecasting systems. The long-range displacement term used in DT also
indicates that cases with longer lead times or larger spread benefit most from
this method. A plausible explanation is that a small number of distant
trajectories can strongly displace the arithmetic ensemble mean, particularly
when distinct track branches develop \cite{tsai2013detection}. Removing these
geometrically anomalous members reduces their influence. However, DT provides
little improvement for low- and medium-spread cases, and a distant branch may
occasionally represent a plausible alternative rather than an erroneous
forecast. Large spread can also arise from compensating forecast errors, so
removing the wrong branch can degrade the consensus \cite{elsberry2000consensus}.
DT is therefore best presented as an additional deterministic track, while the
untrimmed ensemble members remain available for probabilistic interpretation.

HIST is useful when persistent differences in forecast skill exist. In the
present application, however, HIST remains close to the full ensemble mean and
does not provide a consistent operational advantage. Unlike separate models in
a multimodel consensus, perturbed members within a single ensemble do not
represent permanently distinct forecasting systems. Their relative performance
can vary with the atmospheric flow \cite{leutbecher2008ensemble}, so averaging
skill by member index can remove much of the case-specific information. These
results suggest that historical guidance may be more effective when combined
with other statistical methods, rather than when members are removed solely on
the basis of past performance. Historical performance may also become more
useful if it is conditioned on model version, basin, lead time, storm motion,
or forecast regime rather than member identity alone.

SLET has potential as a delayed forecast update. Its relatively high
case-level success rate indicates that early observed track errors contain
useful information for identifying poorly evolving members. Its main weakness
is latency: the method cannot be issued with the original forecast because it
requires the observed storm position at or shortly after genesis. Its
improvement also decreases at longer lead times, plausibly because proximity
to the observed storm early in the forecast does not ensure that a member
correctly represents the subsequent steering flow. Operational testing must
also replace finalized best-track positions with the preliminary storm
positions available in real time.

RC is valuable for identifying and communicating distinct track scenarios. It
can reveal alternative evolutions that are obscured by a single ensemble mean.
The relationship between small cluster size and larger average error motivates
cluster pruning \cite{kowaleski2020use}, but this relationship is statistical
and does not guarantee that a small cluster is incorrect for an individual
storm. This helps explain the center-dependent performance and the degradation
found for some IFS forecasts. RC is also the most computationally expensive
method when all EM initializations are included. Operational products should
therefore display the clusters and their populations before using small-cluster
removal to generate a corrected track.

The two ML methods used in this study, CNN and GBT, offer the potential to
learn nonlinear relationships from complete ensemble trajectories and predict
a new track rather than trim existing members. This is a fundamental difference
from trimming methods: trimming can select a better solution already represented
within the ensemble, whereas ML can produce a solution not represented by any
member. The CNN improvements remain modest, suggesting that forecast tracks
alone may not provide sufficient information for ML-based correction. One
possible explanation is that latitude--longitude sequences contain limited
information about the environmental steering processes responsible for track
error. The expanded predictor set used by GBT likely explains its stronger
performance relative to CNN.

The relatively small training sample for each system and initialization time
may further limit generalization. This may also explain why the CNN method did
not achieve the same level of improvement reported by \cite{agrawal2025using},
where samples were drawn from a single ocean basin. Tropical cyclones within
the same basin may share more similar atmospheric conditions, whereas the
dataset used here spans recent major TCs across multiple basins. Larger
archives, explicit missing-member indicators, and environmental predictors may
allow ML models to extract more transferable relationships. A remaining
challenge is instability and overcorrection: the ML methods can degrade
short-range and low-spread forecasts and improve only a majority of individual
cases. At short lead times, the full ensemble mean is already relatively
accurate and the required correction is small; an unconstrained learned
correction can therefore introduce an error larger than the one it removes.
This overcorrection may also result from poor model initialization and
convergence to a local minimum. The stronger performance in high-spread cases
suggests that the correction signal becomes more identifiable when baseline
errors are larger.

\subsection{Limitations and Future Work}

The dataset includes some of the most intense and destructive TCs during
2016--2024 in order to maximize the availability of forecast records at longer
lead times. Consequently, the results may not generalize to weaker,
short-lived, or developing storms, and the representation of basins, seasons,
and forecast regimes is uneven. Evaluation over additional storm classes,
independent future seasons, and individual model-version periods is therefore
needed.

The ML experiments use a restricted predictor set. This design allows the
effects of ensemble-track information to be evaluated consistently across the
three forecasting systems, but it does not represent the full potential of
ML-based postprocessing. Future work could incorporate spatial atmospheric
fields that describe the large-scale steering flow, environmental circulation,
and storm structure, such as multilevel wind, geopotential height, sea-level
pressure, vorticity, and storm-centered fields. Fused deep-learning approaches
that combine historical trajectories with three-dimensional reanalysis fields
\cite{giffard2020tropical,boussioux2022hurricane,cheung2022hybrid} may be
especially useful. Such an extension is beyond the scope of the present
track-only comparison because it requires consistent gridded fields,
preprocessing procedures, and model-version handling across all three ensemble
systems.

As noted above, the ML methods can also suffer from instability and
overcorrection, particularly near initialization. Future implementations should
therefore include stronger physical and statistical constraints, such as
forcing the correction to zero at initialization, increasing the permitted
correction with lead time, penalizing large or temporally irregular
corrections, and scaling correction limits according to ensemble spread.
Separate short-range models and lead-weighted loss functions may also improve
training stability.

Finally, the retrospective results are not directly equivalent to performance
in a real forecasting environment. A prospective real-time trial should
evaluate data latency, automated quality control, missing-member behavior,
tracker failures, computational cost, and failure recovery, while comparing the
resulting guidance with operational consensus forecasts and human forecaster
decisions.

\section{Conclusion}

This study evaluated six objective approaches for improving ensemble TC track
forecasts from GEFS, MOGREPS-G, and IFS during 2016--2024. The results show
that the value of error-reduction methods depends strongly on lead time,
ensemble spread, and forecasting system. No single method
should replace the full ensemble; instead, these approaches are best viewed as
role-specific guidance that complements probabilistic interpretation.

Among the statistical methods, DT provides the most consistent and operationally
practical improvement. It is forecast-only, computationally efficient, and most
beneficial at longer lead times and in high-spread cases, where anomalous
ensemble branches can strongly displace the ensemble mean. SLET can provide a
useful delayed update after early observations become available, although its
performance degrades with lead time. RC is most valuable for identifying and
communicating distinct track scenarios. HIST provides limited benefit when
applied only by member identity, suggesting that historical performance may need
to be conditioned on forecast regime, basin, lead time, or model version. The ML methods show potential, especially when broader predictors are used, and
achieve their best performance at medium-range lead times. However, their
performance is less stable than DT. In particular, unconstrained ML models
can overcorrect short-range and low-spread forecasts, where the full ensemble
mean is already relatively accurate. Future ML-based postprocessing should
therefore incorporate environmental predictors and stronger physical or
statistical constraints on the correction. These results help connect methodological development with operational forecast use by
showing that simple forecast-only approaches can provide stable gains in
high-uncertainty situations, while more flexible ML methods require additional
constraints before they can be used reliably. The framework and findings therefore provide a basis for developing complementary deterministic guidance from ensemble forecasts without replacing the probabilistic information carried by the full ensemble.

\section*{Data Availability}
TC track forecasts used in this study were downloaded from this link: \url{https://gdex.ucar.edu/datasets/d330003/}. The IBTrACS observational data were downloaded from here: \url{https://www.ncei.noaa.gov/products/international-best-track-archive}. The codes used for data processing, training and evaluation of the models can be found at \url{https://github.com/shuochenw/tc}.

\section*{Acknowledgments}
Research was sponsored by the Transformation Decision Analysis Center. The views and conclusions contained in this document are those of the authors and should not be interpreted as representing the official policies, either expressed or implied, of the Transformation Decision Analysis Center or the U.S. Government. The U.S. Government is authorized to reproduce and distribute reprints for Government purposes notwithstanding any copyright notation herein.

\bibliography{tackling_climate_workshop.bib}

\clearpage

\appendix
\setcounter{section}{0}
\section{Supplementary Tables and Figures}
\label{appendix_a}
\setcounter{table}{0}
\renewcommand{\thetable}{\Alph{section}\arabic{table}}
\renewcommand*{\theHtable}{\thetable}
\setcounter{figure}{0}
\renewcommand{\thefigure}{\Alph{section}\arabic{figure}}
\renewcommand*{\theHfigure}{\thefigure}

\begin{table}[htbp]
\centering
\footnotesize
\caption{The 138 distinct tropical cyclones included in the analysis, grouped by year and basin. Each TC is listed once across forecasting centers and initialization times. Basin abbreviations are AL: North Atlantic; EP: Eastern North Pacific; CP: Central North Pacific; WP: Western North Pacific; NI: North Indian Ocean; SI: South Indian Ocean; and SP: South Pacific.}
\label{tab:tc_names_year_basin}
\resizebox{\linewidth}{!}{%
\begin{tabular}{llrl}
\toprule
Year & Basin & $N$ & Name \\
\midrule
\multirow{5}{*}{2016}
 & AL & 2 & Matthew, Nicole \\
 & EP & 5 & Blas, Georgette, Lester, Madeline, Seymour \\
 & WP & 9 & Chaba, Haima, Lionrock, Malakas, Megi, Meranti, Nepartak, Nock-Ten, Sarika \\
 & SI & 3 & Emeraude, Fantala, Uriah \\
 & SP & 2 & Ula, Winston \\
\midrule
\multirow{4}{*}{2017}
 & AL & 3 & Harvey, Irma, Jose \\
 & WP & 1 & Talim \\
 & SI & 1 & Enawo \\
 & SP & 2 & Debbie, Donna \\
\midrule
\multirow{6}{*}{2018}
 & AL & 2 & Florence, Michael \\
 & EP & 8 & Aletta, Bud, Hector, Lane, Norman, Olivia, Rosa, Sergio \\
 & CP & 1 & Walaka \\
 & WP & 7 & Cimaron, Jebi, Jelawat, Mangkhut, Maria, Trami, Yutu \\
 & SI & 5 & Alcide, Cebile, Cilida, Kenanga, Marcus \\
 & SP & 1 & Gita \\
\midrule
\multirow{5}{*}{2019}
 & AL & 2 & Dorian, Lorenzo \\
 & EP & 3 & Barbara, Erick, Kiko \\
 & WP & 9 & Bualoi, Faxai, Fengshen, Hagibis, Halong, Kammuri, Lekima, Lingling, Wutip \\
 & NI & 2 & Fani, Kyarr \\
 & SI & 5 & Gelena, Haleh, Idai, Joaninha, Veronica \\
\midrule
\multirow{4}{*}{2020}
 & AL & 4 & Delta, Eta, Laura, Teddy \\
 & EP & 2 & Douglas, Marie \\
 & WP & 4 & Goni, Haishen, Maysak, Vamco \\
 & SP & 2 & Harold, Yasa \\
\midrule
\multirow{5}{*}{2021}
 & AL & 2 & Ida, Sam \\
 & EP & 2 & Felicia, Linda \\
 & WP & 4 & Chanthu, Mindulle, Rai, Surigae \\
 & SI & 2 & Faraji, Habana \\
 & SP & 1 & Niran \\
\midrule
\multirow{4}{*}{2022}
 & AL & 2 & Fiona, Ian \\
 & EP & 1 & Darby \\
 & WP & 3 & Hinnamnor, Nanmadol, Noru \\
 & SI & 5 & Batsirai, Darian, Emnati, Halima, Vernon \\
\midrule
\multirow{5}{*}{2023}
 & AL & 3 & Franklin, Idalia, Lee \\
 & EP & 4 & Dora, Fernanda, Jova, Lidia \\
 & WP & 7 & Bolaven, Doksuri, Khanun, Koinu, Lan, Mawar, Saola \\
 & SI & 1 & Freddy \\
 & SP & 2 & Jasper, Kevin \\
\midrule
\multirow{4}{*}{2024}
 & AL & 3 & Beryl, Kirk, Milton \\
 & EP & 2 & Gilma, Kristy \\
 & WP & 6 & Gaemi, Kong-Rey, Man-Yi, Shanshan, Yagi, Yinxing \\
 & SI & 3 & Anggrek, Bheki, Chido \\
\midrule
\multicolumn{2}{r}{\textbf{Total}} & \textbf{138} & \\
\bottomrule
\end{tabular}%
}
\end{table}

\begin{table}[htbp]
\centering
\footnotesize
\setlength{\tabcolsep}{4pt}
\renewcommand{\arraystretch}{1.12}
\caption{Predictor groups used by the five-fold GBT model. The compact notation represents all forecast-only predictors implemented in the model. Brace notation denotes every listed combination. The training targets are the eastward and northward corrections from the full ensemble-mean position to the observed TC position. Observations, full-mean error, forecast center, and initialization offset are not predictors; separate models are fitted for each center--offset combination.}
\label{tab:ml_new_track_predictors}
\begin{tabular}{>{\RaggedRight\arraybackslash}p{0.17\linewidth} >{\RaggedRight\arraybackslash}p{0.39\linewidth} >{\RaggedRight\arraybackslash}p{0.36\linewidth}}
\toprule
Category & Variables & Description \\
\midrule
Lead time
& \texttt{lead\_\allowbreak{}time}
& Forecast lead $t$ (h). \\

Ensemble size
& \texttt{n\_\allowbreak{}members}
& Available perturbed-member count. \\

Ensemble position
& \texttt{ens\_\allowbreak{}\{mean,median\}\_\allowbreak{}\{lat,lon\}}; \texttt{ens\_\allowbreak{}mean\_\allowbreak{}\{lat,lon\}\_\allowbreak{}\{sin,cos\}}
& Mean and median ensemble position, plus cyclic sine--cosine encodings of the mean latitude and longitude. \\

Member-offset distribution
& \texttt{member\_\allowbreak{}\{east,north\}\_\allowbreak{}q\{10,25,50,75,90\}\_\allowbreak{}km}
& Five quantiles of member eastward and northward great-circle offsets from the ensemble-mean position. \\

Ensemble spread
& \texttt{ensemble\_\allowbreak{}\{mean,median,p90,max\}\_\allowbreak{}spread\_\allowbreak{}km}; \texttt{ensemble\_\allowbreak{}std\_\allowbreak{}\{lat,lon\}\_\allowbreak{}deg}
& Mean, median, 90th-percentile, and maximum radial member distance, together with latitude and longitude standard deviations. \\

Motion and spread tendency
& \texttt{ensemble\_\allowbreak{}mean\_\allowbreak{}motion\_\allowbreak{}\{east,north,speed\}\_\allowbreak{}kmh}; \texttt{ensemble\_\allowbreak{}spread\_\allowbreak{}change\_\allowbreak{}km}
& Ensemble-mean motion since the preceding lead and the corresponding change in mean ensemble spread. \\

Track displacement
& \texttt{ensemble\_\allowbreak{}mean\_\allowbreak{}displacement\_\allowbreak{}\{east,north\}\_\allowbreak{}km}; \texttt{ensemble\_\allowbreak{}mean\_\allowbreak{}displacement\_\allowbreak{}km}
& Eastward, northward, and total displacement of the ensemble mean from its first available lead. \\

Initialization date
& \texttt{case\_\allowbreak{}init\_\allowbreak{}\{year,month,dayofyear\}} & Initialization year, month and day-of-year encodings. \\

Pressure
& \texttt{ens\_\allowbreak{}\{mean,min,max,std\}\_\allowbreak{}pressure\_\allowbreak{}hPa}
& Mean, minimum, maximum, and standard deviation of member central pressure. \\

Wind
& \texttt{ens\_\allowbreak{}\{mean,min,max,std\}\_\allowbreak{}wind\_\allowbreak{}mps}
& Mean, minimum, maximum, and standard deviation of member maximum wind speed. \\
\bottomrule
\end{tabular}
\end{table}

\begin{figure}[htbp]
    \centering
    \includegraphics[scale=0.4]{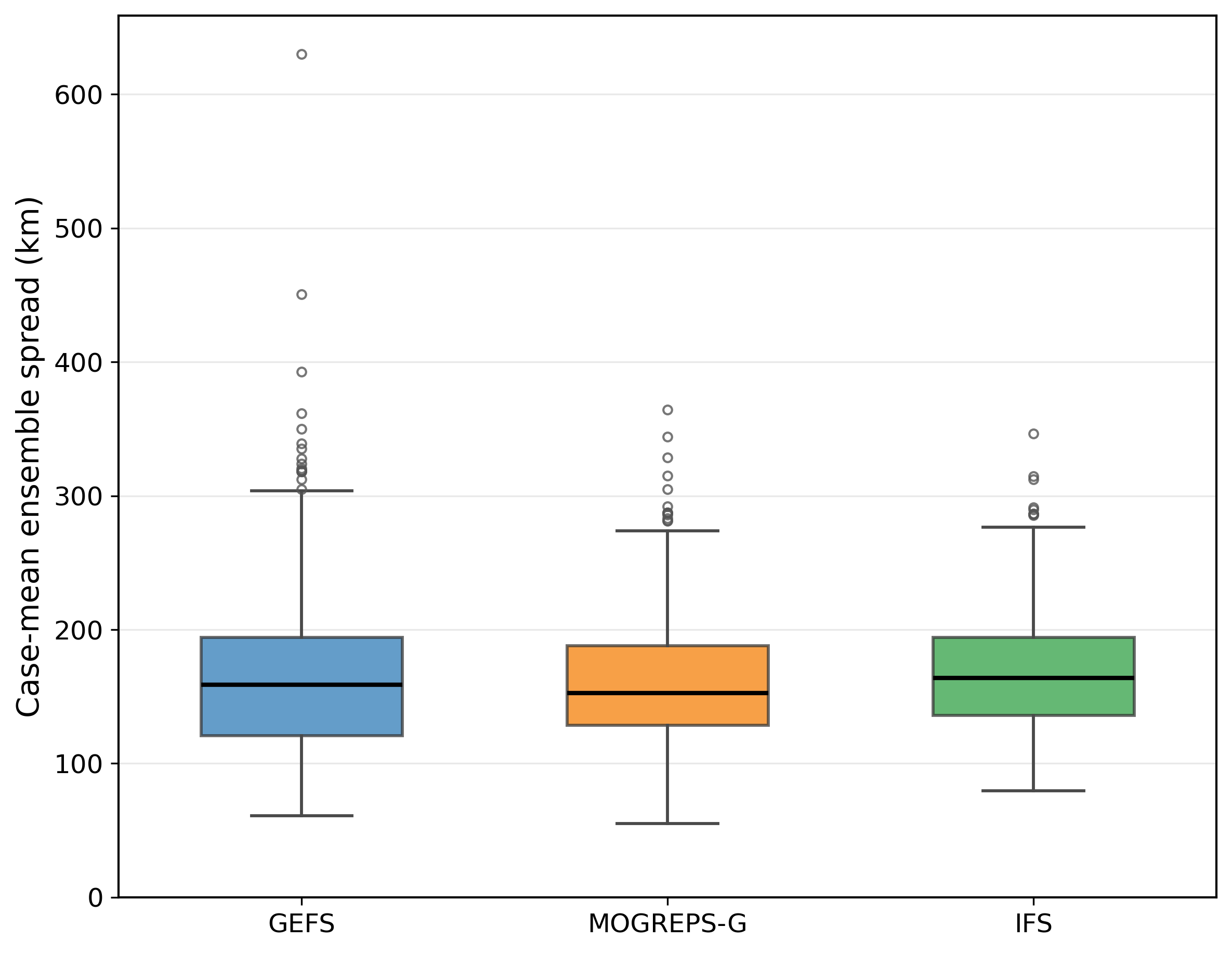}
\caption{Distributions of case-mean ensemble spread for GEFS, MOGREPS-G, and
IFS. At each lead, spread is the mean great-circle distance of the perturbed
members from the full ensemble-mean position; the case mean averages this
quantity over the 25 lead times from 0 to 144 h.}
\label{Fig:ensemble_spread_boxplot_by_center}
\end{figure}

\begin{figure}[t]
\centering
\includegraphics[scale=0.4]
{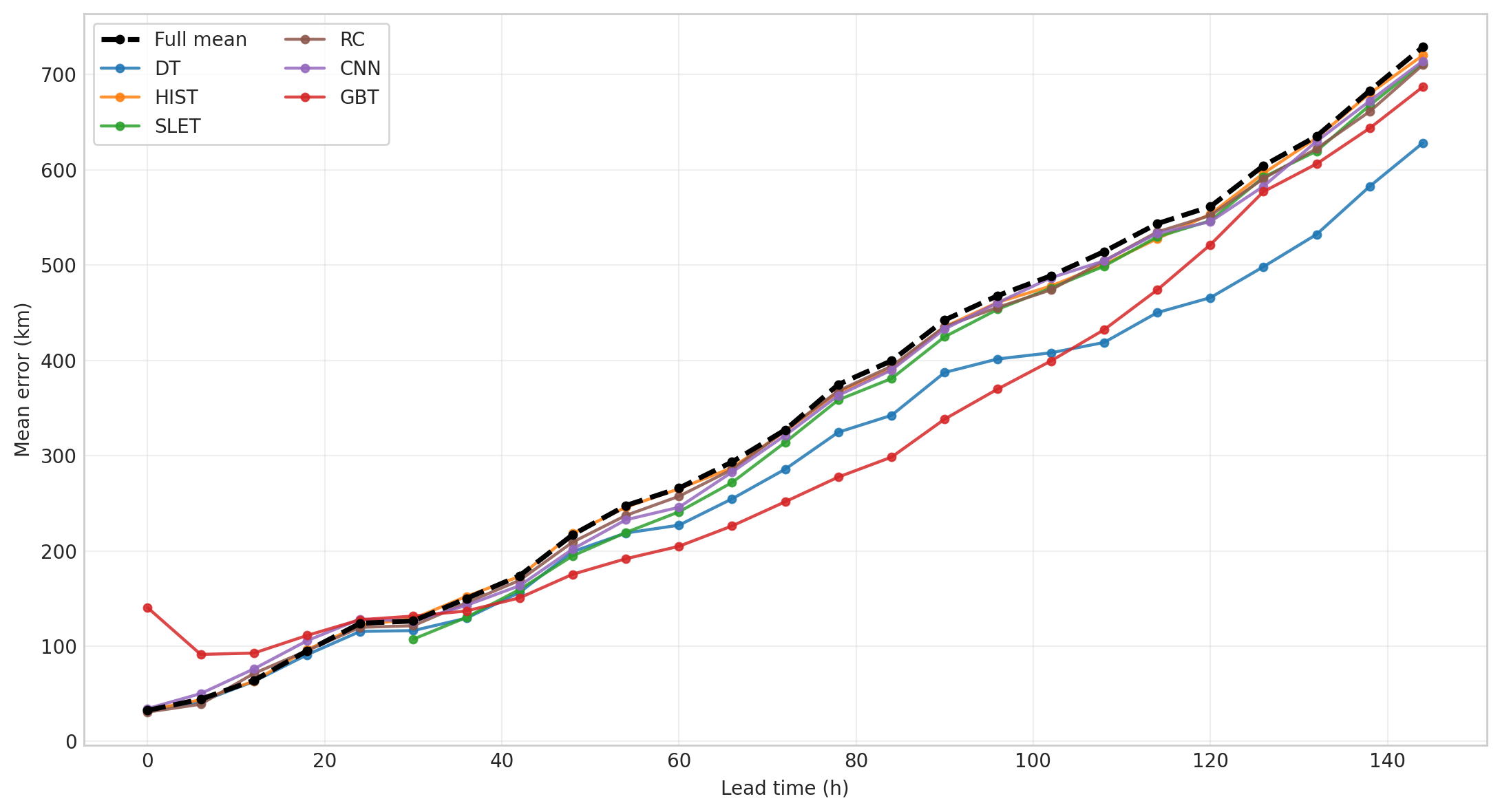}
\caption{Mean total track error as a function of forecast lead time for the
full ensemble mean and the six uncertainty-reduction methods. SLET begins at 30 h since it uses the observed 24-h position for member selection and is assessed only at subsequent lead times. Lower values indicate better performance.}
\label{Fig:method_error_by_lead_all_centers}
\end{figure}

\begin{table}[htbp]
\centering
\footnotesize
\setlength{\tabcolsep}{3pt}
\caption{Track-error improvement relative to the full ensemble mean for GEFS, pooled across 4 forecast initialization times. Entries show percentage improvement with absolute improvement in km in parentheses. Among the six forecast methods, the best method at each lead time is shown in red and the worst in blue. $N$ is the number of matched TC--initialization cases available at each lead time; each TC may contribute up to four initialization times.}
\label{tab:gefs_six_method_improvement}
\resizebox{\linewidth}{!}{%
\begin{tabular}{rrrcccccc}
\toprule
\multicolumn{1}{c}{Lead} & \multirow{2}{*}{$N$} & \multicolumn{1}{c}{Baseline} & \multirow{2}{*}{DT} & \multirow{2}{*}{HIST} & \multirow{2}{*}{SLET} & \multirow{2}{*}{RC} & \multirow{2}{*}{CNN} & \multirow{2}{*}{GBT} \\
\multicolumn{1}{c}{(h)} & & \multicolumn{1}{c}{error (km)} & & & & & & \\
\midrule
0 & 410 & 26.4 & -0.3\% (-0.1) & \textcolor{red}{0.0\% (0.0)} & -- & -1.0\% (-0.3) & -13.4\% (-3.5) & \textcolor{blue}{-419.0\% (-110.4)} \\
6 & 410 & 46.7 & 4.6\% (2.2) & -2.9\% (-1.3) & -- & \textcolor{red}{21.7\% (10.1)} & -15.8\% (-7.3) & \textcolor{blue}{-82.7\% (-38.6)} \\
12 & 410 & 65.9 & \textcolor{red}{8.5\% (5.6)} & 3.9\% (2.6) & -- & -26.7\% (-17.6) & -20.8\% (-13.6) & \textcolor{blue}{-49.6\% (-32.7)} \\
18 & 410 & 90.0 & \textcolor{red}{2.8\% (2.6)} & 1.0\% (0.9) & -- & \textcolor{blue}{-28.1\% (-25.3)} & -13.3\% (-11.9) & -20.8\% (-18.8) \\
24 & 410 & 143.3 & \textcolor{red}{10.0\% (14.3)} & 2.4\% (3.4) & -- & 9.8\% (14.1) & -0.5\% (-0.7) & \textcolor{blue}{-8.0\% (-11.5)} \\
30 & 410 & 138.4 & 13.3\% (18.5) & 0.7\% (1.0) & \textcolor{red}{16.3\% (22.5)} & 8.3\% (11.5) & 1.4\% (2.0) & \textcolor{blue}{-5.7\% (-7.9)} \\
36 & 410 & 178.2 & \textcolor{red}{20.9\% (37.2)} & \textcolor{blue}{0.3\% (0.5)} & 11.6\% (20.7) & 3.0\% (5.3) & 8.1\% (14.4) & 12.1\% (21.6) \\
42 & 410 & 203.4 & 13.1\% (26.5) & 3.2\% (6.5) & 6.0\% (12.2) & \textcolor{blue}{-0.2\% (-0.4)} & 8.1\% (16.4) & \textcolor{red}{17.8\% (36.1)} \\
48 & 410 & 218.7 & 13.9\% (30.4) & 2.5\% (5.5) & 14.0\% (30.5) & \textcolor{blue}{-2.5\% (-5.4)} & 8.9\% (19.5) & \textcolor{red}{14.2\% (31.2)} \\
54 & 410 & 267.6 & 12.8\% (34.3) & 2.9\% (7.8) & 14.5\% (38.8) & \textcolor{blue}{2.6\% (7.0)} & 6.9\% (18.4) & \textcolor{red}{23.9\% (63.9)} \\
60 & 410 & 293.1 & 11.3\% (33.2) & 1.8\% (5.3) & 9.8\% (28.7) & \textcolor{blue}{0.9\% (2.7)} & 6.3\% (18.5) & \textcolor{red}{30.4\% (89.0)} \\
66 & 410 & 302.0 & 10.8\% (32.5) & 5.7\% (17.2) & 6.6\% (19.9) & \textcolor{blue}{2.6\% (8.0)} & 5.0\% (15.1) & \textcolor{red}{21.2\% (63.9)} \\
72 & 410 & 336.6 & 9.4\% (31.6) & 2.6\% (8.7) & 2.5\% (8.3) & \textcolor{blue}{-1.5\% (-5.2)} & 5.2\% (17.4) & \textcolor{red}{25.8\% (86.7)} \\
78 & 410 & 395.6 & 10.3\% (40.6) & 4.0\% (15.7) & 4.2\% (16.6) & \textcolor{blue}{-0.5\% (-2.1)} & 3.8\% (15.1) & \textcolor{red}{34.2\% (135.1)} \\
84 & 410 & 429.6 & 17.3\% (74.5) & 2.1\% (8.9) & 6.2\% (26.4) & \textcolor{blue}{-2.2\% (-9.6)} & 4.4\% (18.7) & \textcolor{red}{26.6\% (114.4)} \\
90 & 410 & 454.4 & 16.7\% (76.1) & 1.5\% (7.0) & 5.4\% (24.5) & \textcolor{blue}{-3.0\% (-13.7)} & 3.3\% (15.1) & \textcolor{red}{24.5\% (111.5)} \\
96 & 410 & 472.9 & 14.1\% (66.7) & \textcolor{blue}{1.3\% (6.4)} & 3.3\% (15.6) & 2.2\% (10.3) & 2.2\% (10.4) & \textcolor{red}{22.4\% (105.8)} \\
102 & 410 & 485.7 & 15.6\% (76.0) & \textcolor{blue}{1.3\% (6.4)} & 3.5\% (17.2) & 4.5\% (21.7) & 2.2\% (10.9) & \textcolor{red}{16.6\% (80.8)} \\
108 & 410 & 506.3 & \textcolor{red}{19.0\% (96.1)} & \textcolor{blue}{1.2\% (6.3)} & 4.4\% (22.2) & 5.6\% (28.4) & 2.0\% (10.2) & 12.4\% (62.8) \\
114 & 410 & 533.7 & \textcolor{red}{19.0\% (101.6)} & \textcolor{blue}{1.3\% (7.0)} & 3.9\% (20.8) & 4.3\% (23.2) & 1.9\% (10.0) & 12.3\% (65.5) \\
120 & 410 & 560.0 & \textcolor{red}{21.3\% (119.5)} & \textcolor{blue}{0.8\% (4.4)} & 3.8\% (21.5) & 6.1\% (34.4) & 1.7\% (9.6) & 9.8\% (55.0) \\
126 & 410 & 611.0 & \textcolor{red}{20.8\% (127.2)} & 0.9\% (5.4) & 3.5\% (21.3) & 5.1\% (31.2) & \textcolor{blue}{0.8\% (4.7)} & 5.4\% (33.1) \\
132 & 410 & 642.0 & \textcolor{red}{18.7\% (120.1)} & \textcolor{blue}{0.8\% (5.2)} & 2.8\% (17.8) & 6.5\% (41.5) & 1.0\% (6.6) & 0.8\% (5.3) \\
138 & 410 & 669.1 & \textcolor{red}{15.8\% (105.5)} & 0.7\% (4.8) & 2.4\% (16.3) & 7.1\% (47.7) & \textcolor{blue}{0.5\% (3.7)} & 3.8\% (25.7) \\
144 & 410 & 684.4 & \textcolor{red}{16.0\% (109.3)} & 2.3\% (15.5) & 3.8\% (25.8) & 6.6\% (45.2) & 0.2\% (1.5) & \textcolor{blue}{-0.2\% (-1.5)} \\
\bottomrule
\end{tabular}%
}
\end{table}

\begin{table}[htbp]
\centering
\footnotesize
\setlength{\tabcolsep}{3pt}
\caption{Same as Table \ref{tab:gefs_six_method_improvement}, but for MOGREPS-G.}
\label{tab:mogreps_g_six_method_improvement}
\resizebox{\linewidth}{!}{%
\begin{tabular}{rrrcccccc}
\toprule
\multicolumn{1}{c}{Lead} & \multirow{2}{*}{$N$} & \multicolumn{1}{c}{Baseline} & \multirow{2}{*}{DT} & \multirow{2}{*}{HIST} & \multirow{2}{*}{SLET} & \multirow{2}{*}{RC} & \multirow{2}{*}{CNN} & \multirow{2}{*}{GBT} \\
\multicolumn{1}{c}{(h)} & & \multicolumn{1}{c}{error (km)} & & & & & & \\
\midrule
0 & 432 & 37.9 & -0.1\% (-0.1) & -0.8\% (-0.3) & -- & \textcolor{red}{8.2\% (3.1)} & -3.4\% (-1.3) & \textcolor{blue}{-201.0\% (-76.3)} \\
6 & 432 & 47.6 & \textcolor{red}{7.6\% (3.6)} & 6.7\% (3.2) & -- & 7.2\% (3.4) & -11.4\% (-5.4) & \textcolor{blue}{-120.4\% (-57.3)} \\
12 & 432 & 81.0 & -2.8\% (-2.3) & \textcolor{red}{-0.3\% (-0.2)} & -- & -5.7\% (-4.6) & -18.2\% (-14.8) & \textcolor{blue}{-25.6\% (-20.7)} \\
18 & 432 & 123.3 & 9.9\% (12.2) & -2.1\% (-2.6) & -- & \textcolor{red}{16.0\% (19.7)} & -8.8\% (-10.9) & \textcolor{blue}{-11.4\% (-14.1)} \\
24 & 432 & 123.8 & \textcolor{red}{7.7\% (9.5)} & 1.2\% (1.5) & -- & -2.5\% (-3.1) & \textcolor{blue}{-5.5\% (-6.8)} & -2.8\% (-3.5) \\
30 & 432 & 131.2 & 1.9\% (2.5) & -4.0\% (-5.2) & \textcolor{red}{18.3\% (24.0)} & 1.5\% (1.9) & -3.4\% (-4.5) & \textcolor{blue}{-10.4\% (-13.7)} \\
36 & 432 & 161.3 & 8.5\% (13.7) & \textcolor{blue}{-2.8\% (-4.5)} & \textcolor{red}{16.7\% (27.0)} & 5.5\% (8.8) & -0.5\% (-0.9) & 8.5\% (13.6) \\
42 & 432 & 183.5 & 6.4\% (11.7) & \textcolor{blue}{-1.5\% (-2.8)} & 10.8\% (19.8) & 8.3\% (15.2) & 1.3\% (2.4) & \textcolor{red}{12.2\% (22.3)} \\
48 & 432 & 247.0 & 4.5\% (11.0) & \textcolor{blue}{-2.2\% (-5.5)} & 10.2\% (25.3) & 13.2\% (32.6) & 6.0\% (14.8) & \textcolor{red}{24.7\% (61.0)} \\
54 & 432 & 278.6 & 7.7\% (21.6) & \textcolor{blue}{-0.7\% (-1.9)} & 9.4\% (26.3) & 9.7\% (26.9) & 5.5\% (15.4) & \textcolor{red}{22.5\% (62.8)} \\
60 & 432 & 299.2 & 17.6\% (52.5) & \textcolor{blue}{-1.1\% (-3.2)} & 10.4\% (31.1) & 7.8\% (23.3) & 8.1\% (24.5) & \textcolor{red}{25.3\% (75.7)} \\
66 & 432 & 326.4 & 16.3\% (53.2) & \textcolor{blue}{-0.1\% (-0.4)} & 10.1\% (33.1) & 4.8\% (15.7) & 8.2\% (26.7) & \textcolor{red}{29.2\% (95.4)} \\
72 & 432 & 386.2 & 15.8\% (61.0) & \textcolor{blue}{-1.5\% (-5.6)} & 6.1\% (23.5) & 2.7\% (10.5) & 2.1\% (8.2) & \textcolor{red}{31.9\% (123.3)} \\
78 & 432 & 430.8 & 15.1\% (65.1) & \textcolor{blue}{0.8\% (3.3)} & 5.5\% (23.9) & 4.8\% (20.5) & 6.4\% (27.7) & \textcolor{red}{28.4\% (122.3)} \\
84 & 432 & 428.4 & 15.2\% (65.2) & \textcolor{blue}{2.9\% (12.6)} & 5.2\% (22.1) & 6.0\% (25.8) & 4.2\% (18.1) & \textcolor{red}{29.5\% (126.5)} \\
90 & 432 & 474.5 & 12.4\% (59.0) & \textcolor{blue}{2.5\% (11.7)} & 4.3\% (20.4) & 7.3\% (34.5) & 4.4\% (20.8) & \textcolor{red}{29.3\% (138.9)} \\
96 & 432 & 512.9 & 16.1\% (82.8) & \textcolor{blue}{2.0\% (10.4)} & 2.5\% (13.0) & 8.7\% (44.5) & 4.9\% (25.3) & \textcolor{red}{26.3\% (134.6)} \\
102 & 432 & 532.8 & 16.9\% (89.9) & 3.1\% (16.3) & \textcolor{blue}{1.6\% (8.6)} & 8.7\% (46.2) & 5.2\% (27.8) & \textcolor{red}{28.0\% (149.1)} \\
108 & 432 & 529.7 & 18.0\% (95.6) & 3.9\% (20.4) & \textcolor{blue}{2.8\% (15.1)} & 6.8\% (35.9) & 4.5\% (23.8) & \textcolor{red}{24.8\% (131.6)} \\
114 & 432 & 550.6 & \textcolor{red}{18.5\% (102.0)} & 4.5\% (25.0) & \textcolor{blue}{2.3\% (12.4)} & 6.6\% (36.6) & 4.6\% (25.5) & 18.2\% (100.1) \\
120 & 432 & 557.1 & \textcolor{red}{14.8\% (82.2)} & \textcolor{blue}{3.0\% (17.0)} & 3.4\% (18.8) & 5.2\% (29.1) & 4.5\% (25.1) & 11.2\% (62.6) \\
126 & 432 & 592.3 & \textcolor{red}{14.1\% (83.8)} & 3.2\% (19.0) & \textcolor{blue}{2.0\% (12.0)} & 5.8\% (34.3) & 4.3\% (25.9) & 10.0\% (59.1) \\
132 & 432 & 635.6 & 13.5\% (85.9) & \textcolor{blue}{-0.1\% (-0.8)} & 4.1\% (26.0) & 3.5\% (22.2) & 2.0\% (13.0) & \textcolor{red}{14.2\% (90.3)} \\
138 & 432 & 711.7 & 14.0\% (99.3) & \textcolor{blue}{0.1\% (0.5)} & 3.6\% (25.7) & 5.6\% (40.1) & 1.1\% (8.0) & \textcolor{red}{15.7\% (112.0)} \\
144 & 432 & 768.0 & 12.2\% (93.8) & \textcolor{blue}{0.3\% (2.5)} & 3.1\% (24.1) & 4.2\% (32.3) & 0.9\% (6.6) & \textcolor{red}{14.8\% (114.0)} \\
\bottomrule
\end{tabular}%
}
\end{table}

\begin{table}[htbp]
\centering
\footnotesize
\setlength{\tabcolsep}{3pt}
\caption{Same as Table \ref{tab:gefs_six_method_improvement}, but for IFS.}
\label{tab:ifs_six_method_improvement}
\resizebox{\linewidth}{!}{%
\begin{tabular}{rrrcccccc}
\toprule
\multicolumn{1}{c}{Lead} & \multirow{2}{*}{$N$} & \multicolumn{1}{c}{Baseline} & \multirow{2}{*}{DT} & \multirow{2}{*}{HIST} & \multirow{2}{*}{SLET} & \multirow{2}{*}{RC} & \multirow{2}{*}{CNN} & \multirow{2}{*}{GBT} \\
\multicolumn{1}{c}{(h)} & & \multicolumn{1}{c}{error (km)} & & & & & & \\
\midrule
0 & 339 & 33.2 & -0.1\% (0.0) & 0.3\% (0.1) & -- & \textcolor{red}{8.3\% (2.8)} & -2.5\% (-0.8) & \textcolor{blue}{-436.3\% (-145.1)} \\
6 & 339 & 37.2 & 4.3\% (1.6) & -0.8\% (-0.3) & -- & \textcolor{red}{4.7\% (1.7)} & -13.2\% (-4.9) & \textcolor{blue}{-118.8\% (-44.1)} \\
12 & 339 & 40.1 & -0.1\% (-0.1) & -0.1\% (0.0) & -- & \textcolor{red}{4.5\% (1.8)} & -15.2\% (-6.1) & \textcolor{blue}{-84.8\% (-34.0)} \\
18 & 339 & 64.9 & -6.8\% (-4.4) & -2.9\% (-1.9) & -- & \textcolor{red}{7.1\% (4.6)} & -11.7\% (-7.6) & \textcolor{blue}{-24.4\% (-15.8)} \\
24 & 339 & 100.8 & 0.1\% (0.1) & 1.3\% (1.3) & -- & 1.5\% (1.5) & \textcolor{blue}{-4.3\% (-4.4)} & \textcolor{red}{5.3\% (5.3)} \\
30 & 339 & 105.8 & \textcolor{red}{9.2\% (9.7)} & \textcolor{blue}{-2.8\% (-3.0)} & 8.2\% (8.7) & 1.0\% (1.1) & 5.1\% (5.5) & 8.5\% (9.0) \\
36 & 339 & 101.4 & 9.3\% (9.4) & \textcolor{blue}{-2.7\% (-2.7)} & 9.6\% (9.7) & -1.9\% (-1.9) & \textcolor{red}{10.4\% (10.6)} & 2.0\% (2.0) \\
42 & 339 & 126.0 & 11.2\% (14.1) & -1.5\% (-1.9) & 8.3\% (10.4) & \textcolor{blue}{-1.5\% (-1.9)} & \textcolor{red}{12.9\% (16.4)} & 7.0\% (8.8) \\
48 & 339 & 176.4 & 6.3\% (11.1) & -2.2\% (-3.9) & 4.5\% (8.0) & \textcolor{blue}{-4.6\% (-8.2)} & 8.0\% (14.2) & \textcolor{red}{16.5\% (29.0)} \\
54 & 339 & 183.7 & 17.1\% (31.5) & -0.5\% (-1.0) & 9.9\% (18.2) & \textcolor{blue}{-3.6\% (-6.6)} & 8.3\% (15.5) & \textcolor{red}{20.2\% (37.2)} \\
60 & 339 & 190.3 & \textcolor{red}{14.8\% (28.2)} & -0.1\% (-0.1) & 6.5\% (12.4) & \textcolor{blue}{-1.7\% (-3.2)} & 11.2\% (21.6) & 4.5\% (8.5) \\
66 & 339 & 240.6 & 11.5\% (27.6) & 0.4\% (1.0) & 3.7\% (8.8) & -0.9\% (-2.1) & \textcolor{blue}{-4.2\% (-10.3)} & \textcolor{red}{14.5\% (35.0)} \\
72 & 339 & 240.3 & \textcolor{red}{11.6\% (27.9)} & 0.8\% (1.9) & 2.3\% (5.6) & -1.8\% (-4.3) & \textcolor{blue}{-2.2\% (-5.3)} & 0.4\% (1.1) \\
78 & 339 & 278.0 & \textcolor{red}{15.3\% (42.5)} & 1.7\% (4.8) & 2.1\% (5.9) & 0.0\% (-0.1) & \textcolor{blue}{-2.7\% (-7.4)} & 7.0\% (19.4) \\
84 & 339 & 326.3 & 8.1\% (26.4) & 0.8\% (2.7) & 1.5\% (4.8) & -0.5\% (-1.5) & \textcolor{blue}{-2.0\% (-6.7)} & \textcolor{red}{16.2\% (52.8)} \\
90 & 339 & 386.8 & 6.3\% (24.5) & 0.4\% (1.4) & 1.5\% (5.8) & -0.7\% (-2.8) & \textcolor{blue}{-1.3\% (-5.2)} & \textcolor{red}{13.1\% (50.5)} \\
96 & 339 & 404.8 & \textcolor{red}{11.2\% (45.5)} & 0.9\% (3.8) & 3.6\% (14.6) & \textcolor{blue}{-6.4\% (-26.0)} & -2.8\% (-11.6) & 10.4\% (42.0) \\
102 & 339 & 436.3 & \textcolor{red}{17.2\% (75.2)} & 2.1\% (9.0) & 3.0\% (13.1) & -7.5\% (-32.9) & \textcolor{blue}{-7.7\% (-33.8)} & 5.4\% (23.7) \\
108 & 339 & 504.0 & \textcolor{red}{18.7\% (94.3)} & 2.7\% (13.5) & 1.4\% (7.2) & \textcolor{blue}{-9.0\% (-45.2)} & -0.1\% (-0.6) & 8.4\% (42.1) \\
114 & 339 & 546.4 & \textcolor{red}{13.3\% (72.5)} & 2.7\% (14.7) & 1.4\% (7.4) & \textcolor{blue}{-7.9\% (-43.3)} & -0.4\% (-2.1) & 6.5\% (35.5) \\
120 & 339 & 568.7 & \textcolor{red}{14.7\% (83.7)} & 0.1\% (0.7) & 0.3\% (1.7) & \textcolor{blue}{-7.9\% (-45.1)} & -0.3\% (-1.8) & -1.2\% (-6.7) \\
126 & 339 & 611.7 & \textcolor{red}{14.6\% (85.8)} & -0.4\% (-2.7) & 0.2\% (1.0) & \textcolor{blue}{-5.4\% (-33.2)} & 3.6\% (21.5) & -3.3\% (-20.4) \\
132 & 339 & 627.0 & \textcolor{red}{16.1\% (100.8)} & 0.2\% (1.2) & -0.3\% (-2.1) & \textcolor{blue}{-5.1\% (-32.0)} & -0.1\% (-0.5) & -3.2\% (-20.4) \\
138 & 339 & 663.6 & \textcolor{red}{14.0\% (92.5)} & 0.7\% (4.7) & -0.2\% (-1.2) & -4.8\% (-31.9) & 4.0\% (26.6) & \textcolor{blue}{-5.6\% (-36.9)} \\
144 & 339 & 733.1 & \textcolor{red}{13.3\% (97.4)} & 1.3\% (9.4) & -0.6\% (-4.7) & \textcolor{blue}{-4.1\% (-30.1)} & 5.4\% (39.7) & 0.3\% (2.1) \\
\bottomrule
\end{tabular}%
}
\end{table}

\begin{figure}[htbp]
\centering
\includegraphics[width=\textwidth]{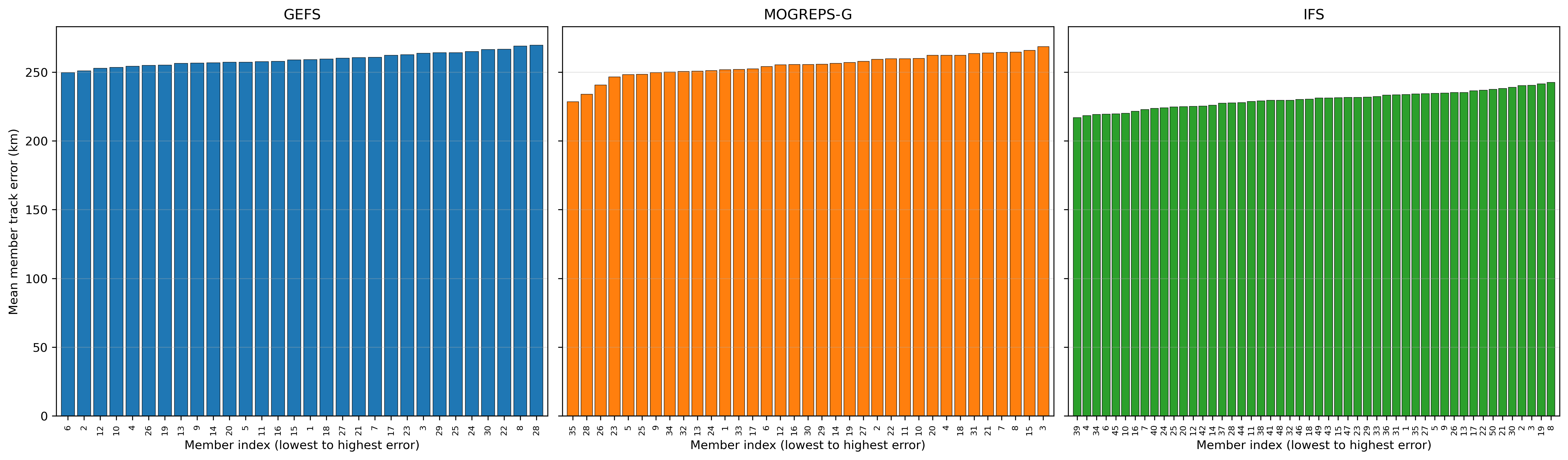}
\caption{Mean track error of individual perturbed ensemble members for GEFS, MOGREPS-G, and IFS. For each TC--initialization case, member error is averaged across all available 0--144 h lead times and then averaged across cases. Within each forecasting system, the actual member indices are ordered from the lowest to the highest mean error.}
\label{fig:member_error_ranking}
\end{figure}

\appendix
\setcounter{section}{1}
\section{Sensitivity Analysis}
\label{appendix_b}
\setcounter{table}{0}
\renewcommand{\thetable}{\Alph{section}\arabic{table}}
\renewcommand*{\theHtable}{\thetable}
\setcounter{figure}{0}
\renewcommand{\thefigure}{\Alph{section}\arabic{figure}}
\renewcommand*{\theHfigure}{\thefigure}

Sensitivity experiments were conducted to evaluate the robustness of the
methods to their principal hyperparameters and to determine the configurations
used in the final analysis. Each hyperparameter was varied individually while
the remaining settings were held at their reference values. For each experiment, performance was measured relative to the full ensemble
mean using
\begin{equation}
I^{(a)}=
100\left(
\frac{\overline{e}^{\mathrm{Full}}-\overline{e}^{(a)}}
{\overline{e}^{\mathrm{Full}}}
\right),
\label{eq:sensitivity_improvement}
\end{equation}
where $\overline{e}^{\mathrm{Full}}$ and $\overline{e}^{(a)}$ are the mean
great-circle track errors of the full ensemble mean and method $a$,
respectively. Both errors were calculated from identical case--lead pairs so
that differences in data availability could not affect the comparison.
Positive values of $I^{(a)}$ indicate improvement over the full ensemble mean.

\begin{table*}[htbp]
\centering
\caption{Hyperparameter search spaces and selected values. Each parameter was
varied individually while the other parameters were held at their reference
values.}
\label{tab:sensitivity_hyperparameters}
\resizebox{\textwidth}{!}{%
\begin{tabular}{lllc}
\toprule
Method & Hyperparameter & Search space & Selected \\

\midrule

\multirow{6}{*}{RC}
& Number of clusters
& $\{3,4,5,6,7\}$ & 7 \\

& Polynomial degree
& $\{1,2,3\}$ & 2 \\

& EM initializations
& $\{10,25,50,100\}$ & 50 \\

& Maximum EM iterations
& $\{100,200,300,500\}$ & 200 \\

& EM convergence tolerance
& $\{10^{-4},10^{-5},10^{-6},10^{-7}\}$ & $10^{-6}$ \\

& Covariance regularization
& $\{10^{-6},10^{-5},10^{-4},10^{-3},10^{-2}\}$ & $10^{-4}$ \\

\midrule

\multirow{4}{*}{CNN}
& Learning rate
& $\{10^{-5},3\!\times\!10^{-5},10^{-4},3\!\times\!10^{-4},
10^{-3},3\!\times\!10^{-3},10^{-2}\}$ & $10^{-3}$ \\

& Convolutional channels
& $\{(16,16),(32,32),(64,32),(64,64),(128,64),(128,128)\}$
& $(64,64)$ \\

& MLP hidden units
& $\{(128,64,32),(256,128,64),(512,256,128),(1024,512,256)\}$
& $(512,256,128)$ \\

& Batch size
& $\{4,8,16,32,64\}$ & 16 \\

\midrule

\multirow{6}{*}{GBT}
& Maximum boosting iterations
& $\{100,150,200,250,300,350,400\}$ & 300 \\

& Learning rate
& $\{0.02,0.03,0.04,0.05,0.06,0.08,0.10\}$ & 0.04 \\

& Maximum leaf nodes
& $\{5,7,10,15,20,25,31\}$ & 7 \\

& Minimum samples per leaf
& $\{10,15,20,25,30,35,40\}$ & 10 \\

& $L_2$ regularization
& $\{0,0.25,0.5,1,2,3,5\}$ & 2 \\

& Batch size
& $\{4,8,16,32,64\}$ & 16 \\

\bottomrule
\end{tabular}%
}
\end{table*}

Table~\ref{tab:sensitivity_hyperparameters} summarizes the candidate values examined for RC, CNN, and GBT and the configurations retained for the final analysis. For CNN training, predictor standardization was estimated from the fitting sample only. Ten percent of the fitting cases was reserved for internal validation. The network was optimized with Adam using mean great-circle track error as the loss, and the gradient norm was limited to 5. Training stopped
when validation error failed to improve. For
GBT, missing predictor values were median-imputed using the fitting sample.
Separate histogram-based boosted-tree models predicted the eastward and
northward corrections. DT and RC do not estimate a supervised mapping from predictors to best-track
positions. DT ranks members using only the geometry of the forecast ensemble,
whereas RC estimates its regression mixture from the forecast trajectories.
For each matched storm and initialization, RC pooled the available trajectories
from the three forecasting systems, unwrapped longitude, normalized lead time,
and fitted the polynomial regression mixture by expectation--maximization. The
fit with the largest likelihood among the retained initializations was used,
and clusters containing less than 20\% of the pooled members were removed.
Neither fitting operation used the best track. Their candidate hyperparameters
were nevertheless evaluated against the best track using the same matched
cases and verification metric as the supervised methods. Consequently, the
sensitivity results quantify tuning behavior on the study archive and should
not be interpreted as an additional independent estimate of out-of-sample
performance.

\begin{figure*}[t]
\centering
\resizebox{0.8\textwidth}{!}{%
\includegraphics{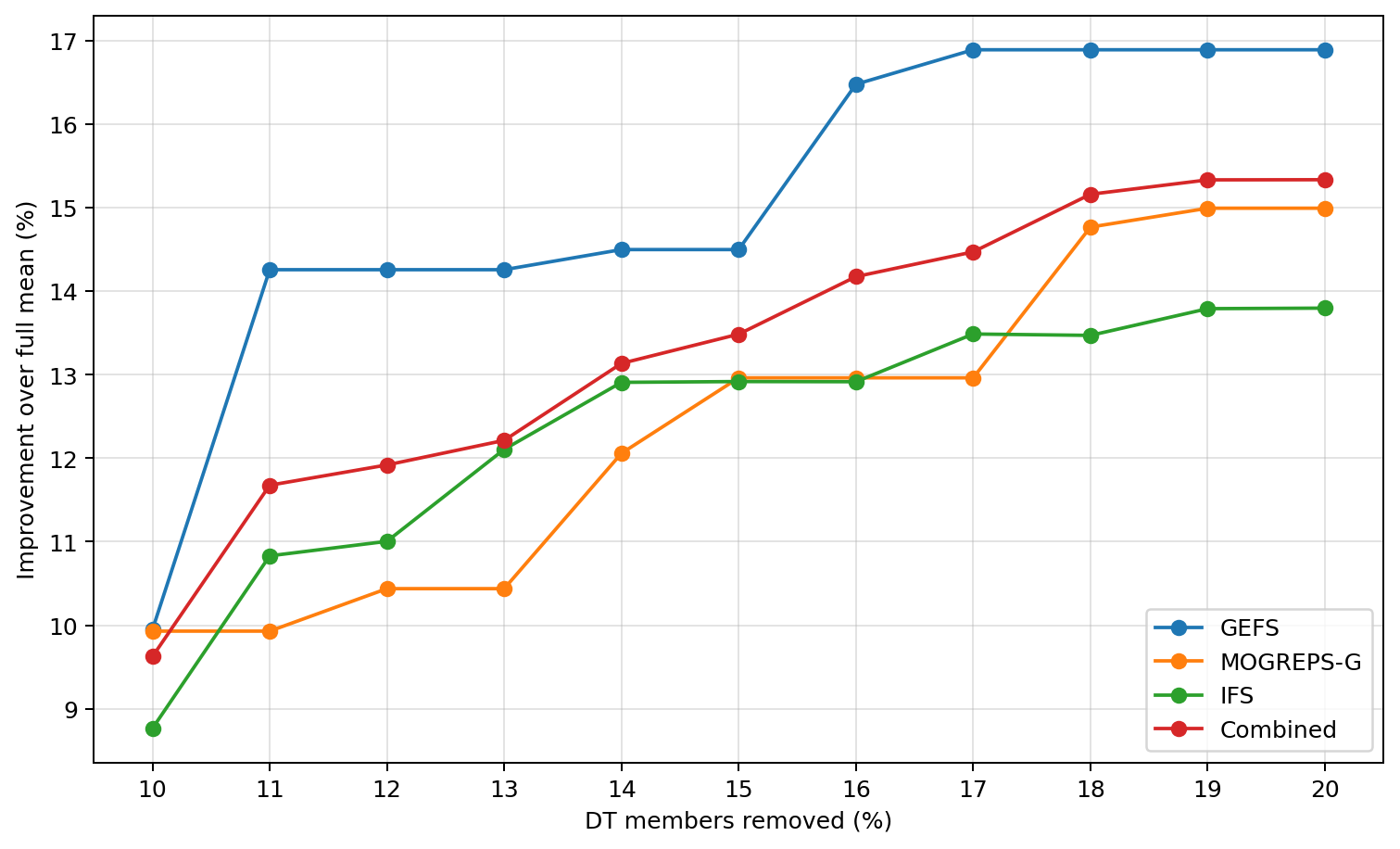}%
}
\caption{Sensitivity of DT performance to the percentage of ensemble members
removed. Improvement relative to the full ensemble mean is calculated over
72--144-h lead times. The combined curve pools all matched case--lead errors
from the three forecasting systems.}
\label{Fig:sensitivity_dt_removal_sensitivity_medium_long}
\end{figure*}

Figure~\ref{Fig:sensitivity_dt_removal_sensitivity_medium_long} shows that DT performance generally improved as the removal fraction increased from 10\% to 20\%. The combined 72--144-h improvement increased from 9.62\% to 15.33\%, with positive responses from all three forecasting systems. Performance changed little between 19\% and 20\%, indicating that the benefit had begun to stabilize. Therefore, 20\% was selected and applied consistently to the other trimming methods. The step-like curves occur because similar percentages can correspond to the same integer number of removed members.

\begin{figure*}[t]
\centering
\resizebox{0.8\textwidth}{!}{%
\includegraphics{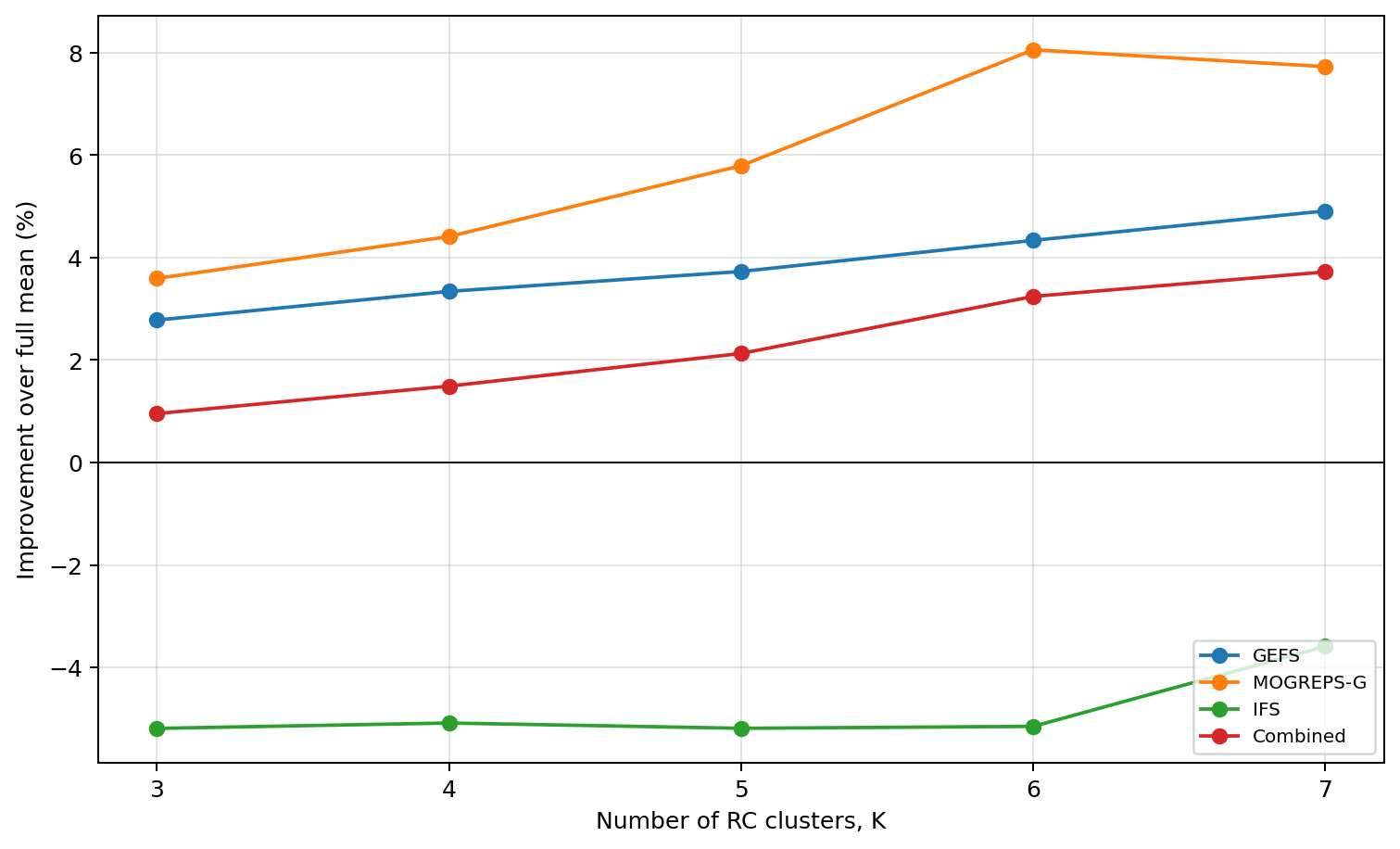}%
}
\caption{Sensitivity of RC performance to the number of regression-mixture
clusters. Improvement relative to the full ensemble mean is calculated over
72--144-h lead times.}
\label{Fig:sensitivity_rc_cluster_sensitivity}
\end{figure*}

\begin{figure*}[t]
\centering
\includegraphics[width=\textwidth]
{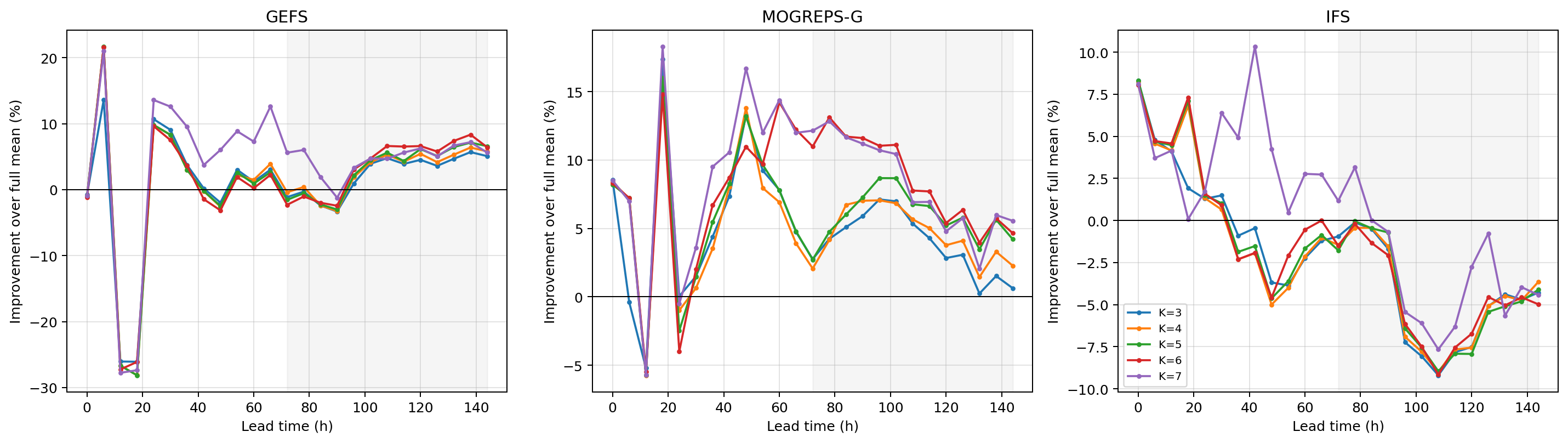}
\caption{RC improvement relative to the full ensemble mean as a function of
forecast lead time for GEFS, MOGREPS-G, and IFS. Curves show candidate cluster
counts from $K=3$ to $K=7$; the shaded region denotes the 72--144-h interval
used for aggregate hyperparameter selection.}
\label{Fig:sensitivity_rc_cluster_sensitivity_by_lead}
\end{figure*}

In Figure~\ref{Fig:sensitivity_rc_cluster_sensitivity}, the combined 72--144-h improvement increased from 0.95\% at $K=3$ to 3.72\% at $K=7$. GEFS improved steadily with increasing $K$, while MOGREPS-G reached its maximum at $K=6$ and remained similar at $K=7$. RC degraded IFS forecasts for every tested value, although the degradation was smallest at $K=7$. Accordingly, $K=7$ was selected because it produced the largest combined improvement. The lead-time results in Figure~\ref{Fig:sensitivity_rc_cluster_sensitivity_by_lead} show substantial short-lead variability. RC generally improved GEFS and MOGREPS-G at medium and long lead times but degraded IFS, particularly during 96--120 h. Thus, RC performance depends strongly on both forecasting system and lead time. Because $K=7$ was the largest tested value, it represents the best candidate examined rather than a global optimum.

\begin{table}[htbp]
\centering
\footnotesize
\caption{Method-specific computational time for 1,181 TC forecast cases across the three forecasting centers and four initialization offsets. Common data loading and shared bookkeeping are excluded. All computations were completed without GPU acceleration.}
\label{tab:method_computational_time}
\begin{tabular}{lrrrrrr}
\toprule
Method & DT & HIST & SLET & RC\emph{*} & CNN & GBT \\
\midrule
Time (s) & 104.3 & 69.9 & 23.5 & 88.6 & 375.2 & 211.7 \\
\bottomrule
\end{tabular}

\par\smallskip
\footnotesize
\emph{*} The computational time for RC is normalized to one of the 50 EM initializations.
\end{table}

In Table \ref{tab:method_computational_time}, we compare the computational cost of all uncertainty-reduction methods used in this study. The statistical methods are computationally efficient, with SLET requiring only 23.5 s for all cases, followed by HIST and DT. The ML methods are more expensive: GBT required 211.7 s, while CNN is the slowest single-model method at 375.2 s because of its iterative neural-network training. The reported RC time of 88.6 s represents one EM initialization; the complete configuration with more initializations makes RC the most computationally demanding method overall. These values show the trade-off between methodological complexity and computational cost.

\end{document}